\documentclass[aps,prb,twocolumn,amsmath,amssymb,superscriptaddress]{revtex4-1}
\usepackage{color}
\usepackage{braket}
\usepackage{siunitx}
\usepackage[dvipsnames]{xcolor}
\usepackage{graphicx}
\graphicspath{{./},{./figures/}}
\usepackage[colorlinks=true,
                linkcolor=blue,
	        citecolor=blue,
	        urlcolor=blue]{hyperref}
\usepackage{titlesec}
\usepackage{tabularx}

\def\a{{\bf a}}
\def\b{{\bf b}}
\def\c{{\bf c}}
\def\d{{\bf d}}

\renewcommand{\figurename}{Fig.} 
\makeatletter	        
\def\fnum@figure{\textbf{\figurename~\thefigure}}
\makeatother
\newcommand{\titledcaption}[2]{\caption{{\textbf{#1}. #2}}}

\makeatletter	        
 \def\section{%
  \@startsection{section}{1}{\z@}{0.8cm plus1ex minus.2ex}{0.2cm}%
  {%
   \small\sffamily\bfseries\selectfont
   \raggedright
   \parindent\z@
  }%
 }%
  \def\subsection{%
  \@startsection{subsection}{2}{\z@}{0.8cm plus1ex minus.2ex}{0.2cm}%
  {%
   \small\sffamily\bfseries
   \raggedright
   \parindent\z@
  }%
 }%
\makeatother
 
\AtBeginDocument{\usepackage{booktabs}}
\begin{document}

\title{Twist-resilient and robust ferroelectric quantum spin Hall insulators driven by van~der~Waals interactions}

\author{Antimo Marrazzo}
\affiliation{\trieste}
\affiliation{\theos}

\author{Marco Gibertini}
\affiliation{\unimore}
\affiliation{\cnrnano}
\affiliation{\theos}

\newcommand{\theos}{Theory and Simulation of Materials (THEOS), and National Centre for Computational Design and Discovery of Novel Materials (MARVEL), \'Ecole Polytechnique F\'ed\'erale de Lausanne, CH-1015 Lausanne, Switzerland}
\newcommand{\trieste}{Dipartimento di Fisica, Universit\`a di Trieste,  I-34151 Trieste, Italy}
\newcommand{\unimore}{Dipartimento di Scienze Fisiche, Informatiche e Matematiche, University of Modena and Reggio Emilia, I-41125 Modena, Italy}
\newcommand{\cnrnano}{Centro S3, CNR-Istituto Nanoscienze, I-41125 Modena, Italy}

\begin{abstract}
Quantum spin Hall insulators (QSHI) have been proposed to power a number of applications, many of which rely on the possibility to switch on and off the non-trivial topology. Typically this control is achieved through strain or external electric fields, which require energy consumption to be maintained. On the contrary, a non-volatile mechanism would be highly beneficial and could be realized through ferroelectricity if opposite polarization states are associated with different topological phases.
While this is not possible in a single ferroelectric material where the two polarization states are related by inversion, the necessary asymmetry could be introduced  by combining a ferroelectric layer with another two-dimensional (2D) trivial insulator. 
Here, by means of first-principles simulations, not only we propose that this is a promising strategy to engineer non-volatile ferroelectric control of topological order in 2D heterostructures, but also that the effect is robust and can survive up to room temperature, irrespective of the weak van der Waals coupling between the layers. We illustrate the general idea by considering a heterostructure made of a well-known ferroelectric material, In$_2$Se$_3$, and a suitably chosen, easily exfoliable trivial insulator, CuI.  In one polarization state the system is trivial, while it becomes a QSHI with a robust band gap upon polarization reversal.
Remarkably, the topological band gap is mediated by the interlayer hybridization and allows to maximise the effect of intralayer spin-orbit coupling, promoting a robust ferroelectric topological phase that could not exist in monolayer materials and is resilient against relative orientation and lattice matching between the layers.
 
\end{abstract}
\maketitle

\section*{Introduction}
Topological insulators (TIs) are characterized by the presence of surface, edge or hinge states protected by a non-trivial topological invariant~\cite{bernevig_topological_2013,vanderbilt_berry_2018}.
These invariants are integer numbers that represent global properties of the bulk electronic wavefunction and induce boundary effects through the so-called bulk-boundary correspondence~\cite{vanderbilt_berry_2018}. Beyond the fundamental interest for the topological physics, a few potential technological applications of TIs have been proposed, ranging from low-dissipation spintronics~\cite{gilbert_topological_2021} to topological quantum computing~\cite{fu_superconducting_2008,lian_topological_2018}. 
Among TIs, time-reversal invariant two-dimensional (2D) TIs---also known as quantum spin Hall insulators~\cite{kane_quantum_2005,kane_$z_2$_2005,bernevig_quantum_2006} (QSHIs)---are particularly relevant from a device perspective. First, and at variance with all the so-called topological crystalline insulators~\cite{fu_crystalline_2011}, QSHIs require only time-reversal symmetry to be preserved~\cite{bernevig_topological_2013,vanderbilt_berry_2018} while being, at the same time, much more abundant than Chern (a.k.a. quantum anomalous Hall) insulators~\cite{vanderbilt_berry_2018,marrazzo_relative_2019,olsen_discovering_2019,vergniory_complete_2019,xu_high-throughput_2020}. Second, QSHIs exhibit one-dimensional (1D) edge states where elastic backscattering is strictly forbidden~\cite{vanderbilt_berry_2018} leading to low-dissipation transport, while in three-dimensional TIs scattering is forbidden only at $\pi$ angles and it is allowed at any other angle. This means that nanoribbons of QSHIs can host 1D low-dissipation wires to be used for nanoelectronics, such as interconnects~\cite{gilbert_topological_2021}. In addition, the spin-momentum locking of the edge states could be exploited in spintronic devices such as spin-current generators and charge-to-spin convertors~\cite{han_quantum_2018}. Finally, QSHIs can leverage the tunability due to their low dimensionality to be manipulated in several ways, ranging from electrical gating to functionalization~\cite{xu_large-gap_2013}, to substrate effects~\cite{reis_bismuthene_2017}, to strain~\cite{huang_bending_2017}.

\begin{figure*}
    \includegraphics[width=0.9\textwidth]{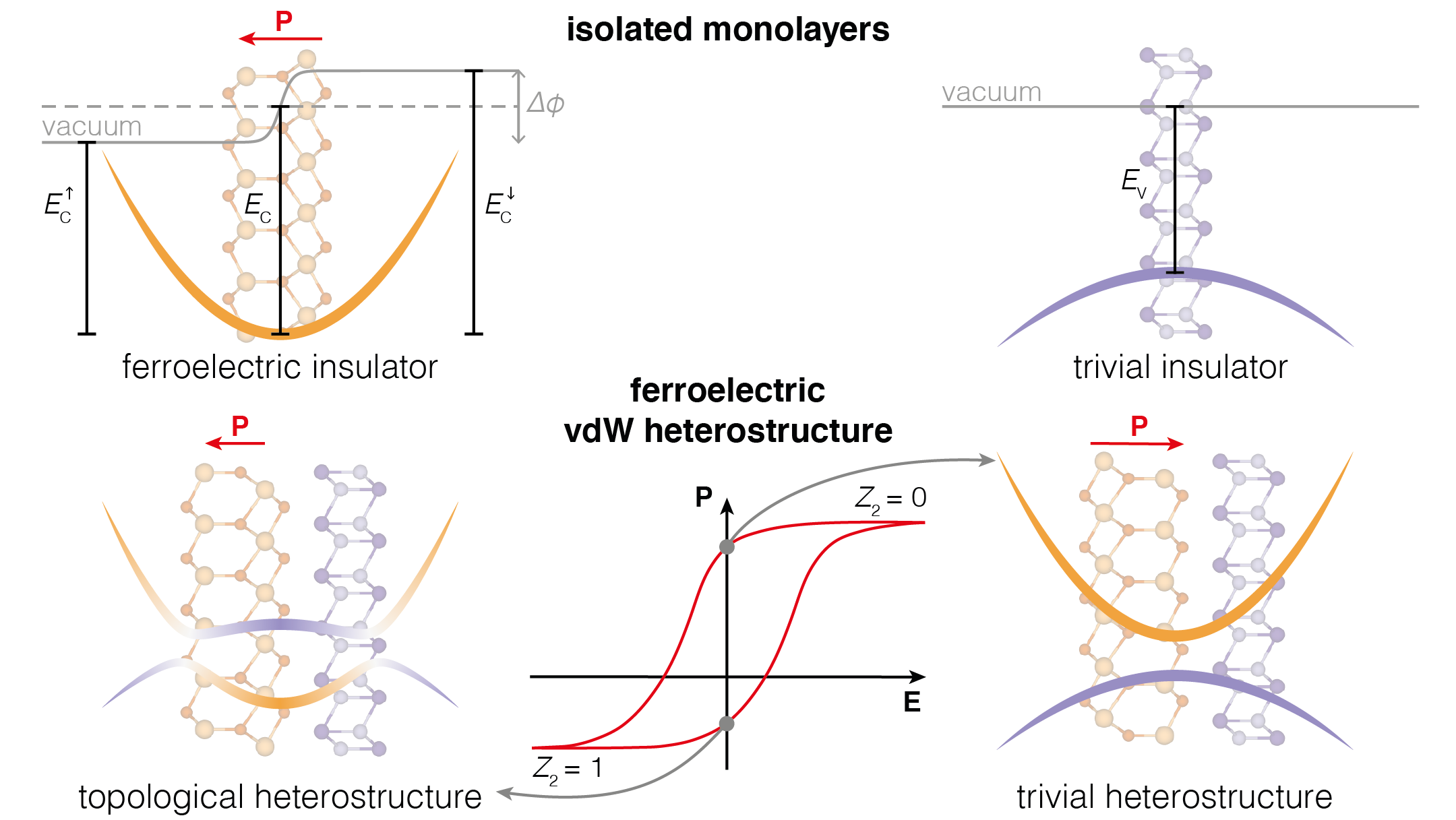}
    \titledcaption{Operating principles of ferroelectric control of topology in a  van-der-Waals heterostructure}{A two-dimensional (2D) ferroelectric insulator with finite out-of-plane polarization and a trivial 2D insulator are put together forming a van-der-Waals heterostructure. As a consequence of the finite vertical dipole,  an electrostatic potential drop $\Delta\phi$ exists across the ferroelectric layer, which shifts the vacuum level. Remarkably, it is possible to find combinations of materials for which, upon the formation of the heterostructure, this difference in vacuum energy can give rise to distinct band alignments between the two layers depending on the polarization direction. In one polarization state the dipole-induced offset is such that a finite energy gap is present between the valence bands associated with the normal layer and the conduction bands arising from the ferroelectric material, so that the system is a trivial semiconductor. For the opposite polarization, a band inversion occurs and spin orbit interactions can open a finite gap between inverted bands, promoting the heterostructure to a quantum spin Hall insulator with a non-trivial $\mathbb{Z}_2$ topological invariant. The difference between the two polarization states is reflected in an asymmetry of the hysteresis cycle, with a smaller magnitude of polarization in the topologically non-trivial state as a consequence of the charge transfer between the layers arising from the band inversion.
}
    \label{fig:scheme}
\end{figure*}

A lot of these applications, such as the topological field-effect transistor (topoFET)~\cite{qian_quantum_2014}, rely on the switching between a topological and a trivial insulating phase driven by an out-of-plane electric field. Typically, edge conductance is turned off for a sufficiently strong gate voltage~\cite{collins_electric-field-tuned_2018} while at zero field the system is a QSHI, although the opposite effect can also be put forward~\cite{liu_switching_2015}.
The transition is typically volatile, meaning that the system goes back to the zero-field state when the gate voltage is removed, thus requiring energy consumption to be maintained.
However, it is of compelling relevance to realize a non-volatile counterpart of this effect, where the material stays in the topological or trivial state even after the field is removed. In this respect, the most prominent way of introducing memory in materials, while preserving  time-reversal symmetry, is through ferroelectricity. Ferroelectric topological transistors would consume energy only to switch and would preserve memory of the state, leading to low-dissipation storage devices and memristors~\cite{chanthbouala_ferroelectric_2012}.
However, the coexistence of ferroelectricity and topological order is rare and often driven by functionalization~\cite{zhao_reversible_2018,kou_two-dimensional_2018} or strain~\cite{liu_strain-induced_2016}.
In addition, in bulk ferroelectric materials the two polarization states are related by inversion symmetry, forcing the topological order to be identical in both states~\cite{narayan_class_2015,liu_strain-induced_2016,monserrat_antiferroelectric_2017}. Similarly, in antiferroelectric topological insulators~\cite{monserrat_antiferroelectric_2017} topological transitions require a finite field to be sustained and would exhibit the same topological or trivial phase for opposite field directions. 
Instead, it would be more relevant for applications to have a ferroelectric structure where opposite polarization states (at zero field) correspond to different topological phases, enabling the non-volatile control of the edge currents.

An interesting perspective is inspired by nature through the easily exfoliable 2D material In$_2$ZnS$_4$~\cite{mounet_two-dimensional_2018} that was recently discovered to be a QSHI by the authors in Refs.~\onlinecite{marrazzo_relative_2019,phd_marrazzo_2019}. A closer inspection to its crystal structure shows that it can be interpreted as a spontaneously occurring van-der-Waals (vdW) heterostructure made of In$_2$S$_3$ and ZrS layers. Taken separately, the two monolayers are semiconducting and topologically trivial, but In$_2$S$_3$ is polar and the band offset associated with the vertical electric dipole drives an inversion between the valence band, associated with one layer, and the conduction band, arising from the other layer, that hybridize with the appearance of a topological gap in the presence of spin-orbit coupling (SOC). 

In this work, we propose that if the polar material is ferroelectric, such vdW heterostructures made of two topologically-trivial 2D materials--namely a trivial insulator and a ferroelectric insulator--behave as a ferroelectric QSHI where the polarization direction and the $\mathbb{Z}_2$ topological invariant are coupled. This happens when valence and conduction bands are associated with different layers and the two polarization states, with opposite offsets stemming from the vertical dipole, give rise to different alignments between them (see Fig.~\ref{fig:scheme}). More specifically, we can have a ferroelectric QSHI if in one polarization state conduction and valence bands are inverted and SOC can open a topological gap, while in the opposite state  the band inversion is suppressed leading to a topologically trivial phase.  Here we show that, not only this is a general strategy to engineer non-volatile ferroelectric control of topological order in 2D heterostructures~\cite{phd_marrazzo_2019,zhang_heterobilayer_2021,bai_nonvolatile_2020,huang_ondemand_2021,liang_intertwined_2021}, but also that the effect is robust and can survive up to room temperature, irrespective of the weak vdW coupling between the layers.
Indeed, we find that, remarkably, when the band inversion occurs at the Brillouin zone (BZ) center, its existence and the associated topological phase are purely driven by band alignment, and thus independent of the relative orientation of the two layers and do not require lattice matching (either in terms of lattice parameters or symmetry). This suggests that, although  stringent conditions on band alignment and sufficiently strong SOC in at least one of the two materials are needed, the range of possible materials combinations is rather large. 
Moreover, we show that, while vdW interactions are notoriously weak and the interlayer distances are typically rather large, the weak interlayer hybridization is fundamental to support robust topological phases driven by band alignment and atomic SOC.

\section*{Results}
\subsection*{Reference system}
Although In$_2$ZnS$_4$ could provide a tantalizing starting point,  it actually displays rather poor performance in terms of band gap~\cite{marrazzo_relative_2019}, for reasons that will be clarified later.
To maximize the effect and illustrate the idea, we search for an optimal combination of monolayers. Ideally, the vdW heterostructure should be made of two easily exfoliable materials with low binding energies~\cite{mounet_two-dimensional_2018} to facilitate fabrication, and display a QSHI phase with a strong band inversion and a relatively large gap to maximize performance. The energy barrier between the two polarization states should also be sufficiently low to be overcome with relatively weak electric fields (of the order of a few tenths of V$/$nm) and sufficiently large to sustain room-temperature ferroelectricity.

In this work, we thus consider In$_2$Se$_3$, a well-known 2D ferroelectric semiconductor~\cite{ding_prediction_2017,zhou_out_2017,xiao_intrinsic_2018,cui_intercorrelated_2018} with the bottom of the conduction band at the BZ center ($\Gamma$ point), and combine it with an optimal semiconducting monolayer from large databases of 2D materials~\cite{mounet_two-dimensional_2018,haastrup_c2db_2018,gjerding_recent_2021}, with a focus on easily exfoliable compounds~\cite{mounet_two-dimensional_2018}. To facilitate simulations, we look for a 2D material that is lattice matched with In$_2$Se$_3$, although this is not crucial for experiments as we shall discuss. More compelling, we require that the top of the valence band is at $\Gamma$ and lies sufficiently close in energy (with respect to vacuum) to the conduction band bottom of In$_2$Se$_3$ (also at $\Gamma$) and that it contains sufficiently heavy elements to be expected to display significant SOC. While these conditions might seem very strict, in reality there are many candidates that can satisfy them according to density-functional theory (DFT) simulations within the PBE approximation~\cite{perdew_pbe_96} (see Supplementary Note 1 and Supplementary Fig.~1). Among them, we find  CuI, an insulator with a 1.8 eV band gap at the DFT-PBE level and the PtTe-prototype structure~\cite{mounet_two-dimensional_2018}, to be optimal for assembling with In$_2$Se$_3$ a vdW ferroelectric QSHI. We note that monolayers of CuI have recently been grown and encapsulated between graphene sheets~\cite{mustonen_towards_2021}.

We stress that this combination of materials is chosen here only for illustrative purposes and that the physics we discuss is very general and it holds for a number of other systems~\cite{wang_tunable_2021} such as In$_2$Se$_3$/PtTe$_2$~\cite{phd_marrazzo_2019} or As~\cite{zhang_heterobilayer_2021}. We thus believe that there is an entire portfolio of prospective heterostructures to be considered  in experimental investigations. In this respect, it is important to bear in mind that the identification of potential candidates in Supplementary Note 1 is based on DFT calculations within routine approximations for the exchange-correlation functional. The accuracy of the calculated band alignment needs thus to be further tested with more sophisticated methods, as approximate DFT tends to underestimate band gaps and work functions. In Supplementary Note 2 we perform such analysis for CuI/In$_2$Se$_3$ (see Supplementary Table 1 for a summary), with a partially positive assessment that this heterostructure could indeed give rise to a ferroelectric QSHI. Similar investigations could be performed also for other prospective systems and would very likely provide an ultimate candidate heterostructure. However, such analysis is computationally very demanding and beyond the illustrative scopes of the current study. 

\subsection*{Electrostatics and band alignment of isolated monolayers}

\begin{figure}
\includegraphics{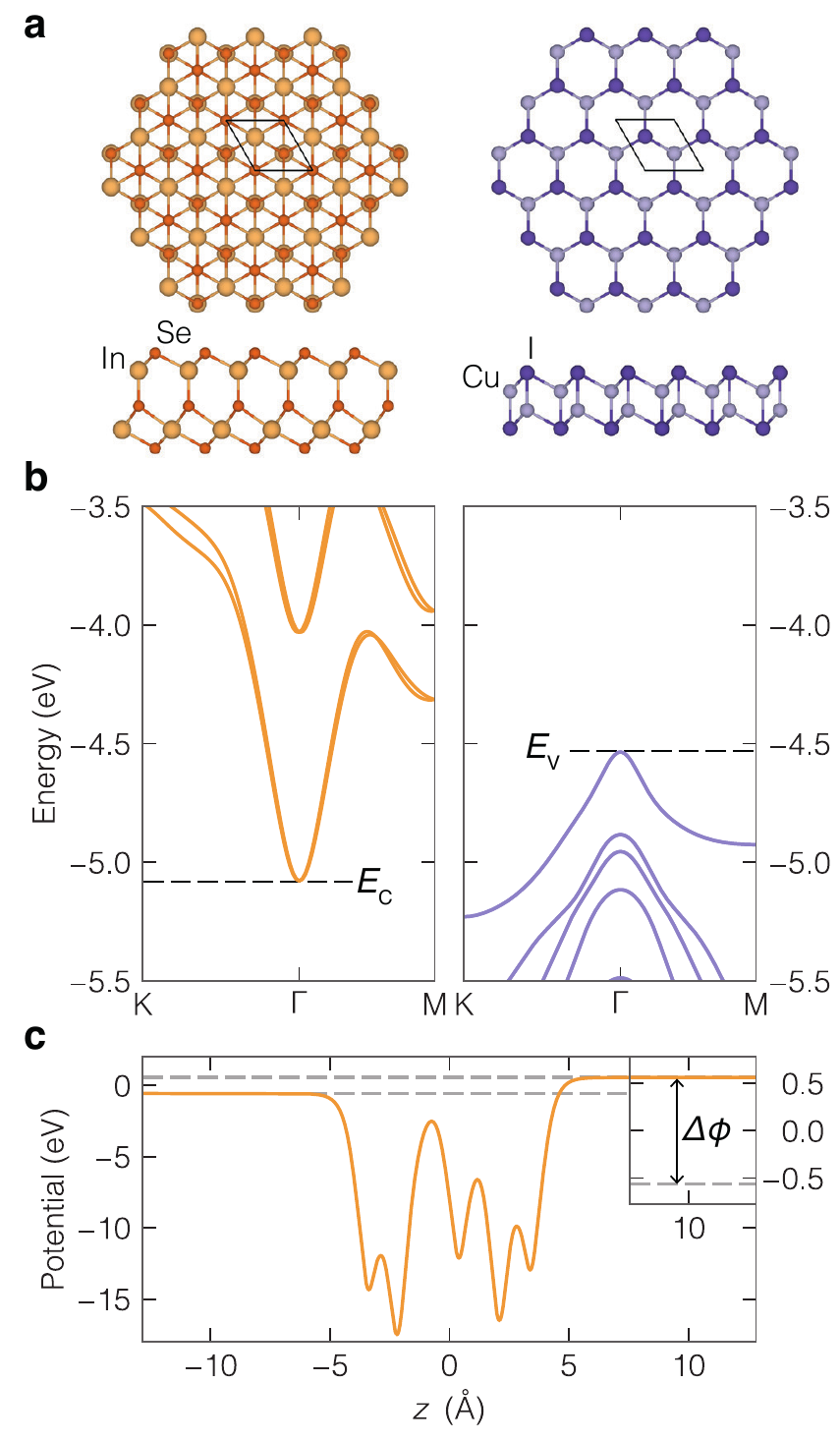}
\titledcaption{Crystal and electronic structure of isolated monolayers}{\a\ Top and lateral views of the crystal structures of In$_2$Se$_3$ (left) and CuI (right). The primitive unit cell is shown with a thin solid line. \b\ Electronic band structure of the two isolated materials, obtained through PBE-DFT calculations including spin-orbit coupling, in a energy range comprising the bottom conduction bands of In$_2$Se$_3$ (orange, left) and the top valence bands of CuI (violet, right). The zero of energy is set at the potential energy of a plane of uniform charge density. \c\ Planar average of the electrostatic potential energy as a function of the vertical coordinate $z$ across a single layer of In$_2$Se$_3$. A clear difference in potential energy $\Delta\phi$ between the two sides is present and its magnitude is emphasized  in the inset. 
\label{fig:isolated}}
\end{figure}

We first report more in detail on the electronic structure of the two isolated monolayers, whose crystal structure is shown in Fig.~\ref{fig:isolated}a.  Both materials have a finite gap separating occupied valence bands from empty conduction bands at zero temperature. In Fig.~\ref{fig:isolated}b we show their energy band dispersion along paths connecting the high-symmetry points K and M to the BZ center $\Gamma$, as obtained through DFT-PBE simulations including SOC, focusing on an energy range where only the conduction bands of In$_2$Se$_3$ and the valence bands of CuI appear. Here the zero of energy is not arbitrary but has well-defined physical meaning associated with the correct open-boundary condition along the vertical direction typical of 2D systems (see Methods for more detail). The conduction band minimum $E_{\rm c}$ of In$_2$Se$_3$ and the valence band maximum $E_{\rm v}$ of CuI both appear at the $\Gamma$ point, with $E_{\rm v}<E_{\rm c}$.

To obtain the correct band alignment when the two materials are sufficiently far away along the vertical direction, we need to take into account the fact that the finite out-of-plane polarization of In$_2$Se$_3$ gives rise to an electrostatic potential energy difference $\Delta\phi$ across the material, as shown in Fig.~\ref{fig:isolated}c with $\Delta\phi\simeq 1$~eV. As a consequence, while in the non-polar CuI the vacuum energy coincides with the zero of energy, in In$_2$Se$_3$ the vacuum energy is shifted by $\mp\Delta\phi/2$ on the two sides of the material, depending on whether the polarization is pointing in that direction or in the opposite. Relative to vacuum, the conduction band minimum then becomes different on the two sides or, equivalently, on a given side for the two polarization states, i.e.\ $E_{\rm c}^{\uparrow,\downarrow} = E_{\rm c} \pm \Delta\phi/2$, as shown schematically in Fig.~\ref{fig:scheme}. When the layers are sufficiently separated, the relative alignment between the energy bands in the two materials can be obtained by equating the corresponding vacuum levels (see Methods) and thus depends on the polarization direction of the ferroelectric layer. When the polarization of In$_2$Se$_3$ is pointing towards CuI, we have that the energy difference between the bottom of the conduction band and the top of the valence band is $\Delta E^{\uparrow} = E^{\uparrow}_{\rm c} - E_{\rm v} = E_{\rm c} - E_{\rm v} + \Delta\phi/2$, while when the polarization points in the opposite direction, away from CuI, we expect $\Delta E^{\downarrow} = E^{\downarrow}_{\rm c} - E_{\rm v} = E_{\rm c} - E_{\rm v} - \Delta\phi/2$. If $\Delta\phi/2  > |E_{\rm c}-E_{\rm v}|$, we can thus have a type II alignment for one polarization state ($\Delta E^{\uparrow}>0$) and a type III alignment for the opposite polarization  ($\Delta E^{\downarrow}<0$). This is the case, although marginally, for In$_2$Se$_3$/CuI, for which $\Delta \phi = 1.12$~eV and $E_{\rm c}-E_{\rm v} = - 0.54$~eV, suggesting that when the polarization points towards CuI there is a finite gap with the bottom of In$_2$Se$_3$ conduction band lying above the top of CuI valence band, while when the polarization points away from CuI there is a band inversion between valence and conduction in the two layers.

\subsection*{Polarization-dependent energy bands of the  heterostructure}

\begin{figure}
   \includegraphics[width=\linewidth]{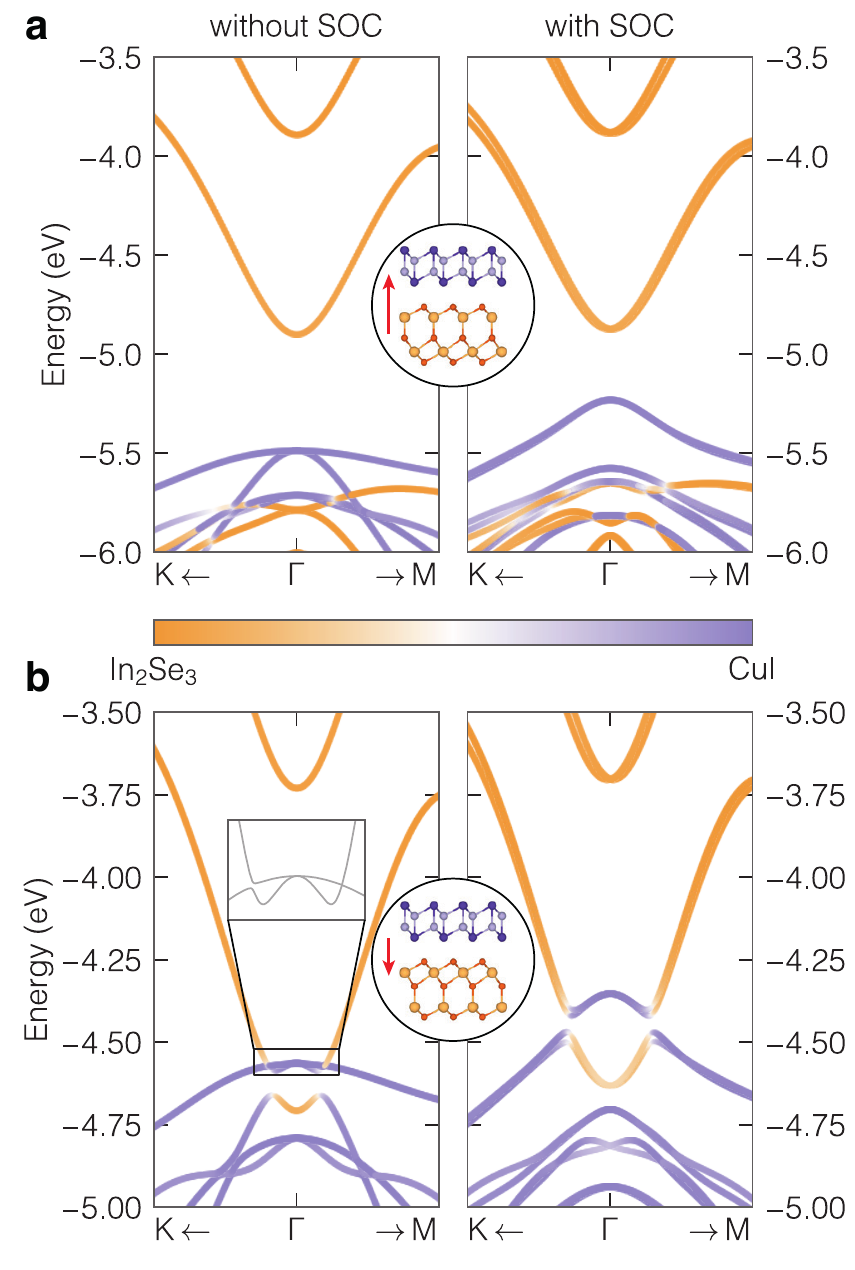}
\titledcaption{Energy bands of In$_2$Se$_3$/CuI heterostructure}{Electronic band structure calculated at the PBE-DFT level  when the polarization points from  In$_2$Se$_3$ to CuI (\a) or when it points in the opposite direction (\b). The color coding represents the layer contribution to electronic states, with orange denoting In$_2$Se$_3$ and violet CuI. Results in left (right) panels have been obtained without (with) the effect of  spin-orbit coupling. The insets show a lateral view of the heterostructure in the most stable configuration, corresponding to a close-packing sequence of the atoms close to the interface. The two polarization states correspond to different band offsets between the monolayers and result, respectively, into a band inversion (\b) or not (\a). In the band-inverted case (\b), the system is metallic without spin-orbit coupling (left), with Dirac crossings between the bands along the $\Gamma$-M direction (see inset), while a full band gap of $52$~meV opens when spin-orbit coupling is included (right), leading to a quantum spin Hall insulating phase.}
    \label{fig:hetero}
\end{figure}

We now want to consider the experimentally relevant case when the two layers are brought at a closer vertical (equilibrium) distance and the band alignment can be affected by possible interface effects, including charge transfer or charge redistribution. Moreover, the hybridization between electronic states in the two layers can introduce subtle effects on the band structure. We thus relax the vdW heterostructure using the rVV10~\cite{vydrov_vv10_09,sabatini_rvv10_13} vdw-compliant functional (more details in the Methods) for the two polarization states and for different horizontal alignments between the layers within the common primitive unit cell. For both polarizations we find that  atoms prefer in-plane high-symmetry positions--with relative coordinates $(0,0)$, $(1/3,2/3)$  or $(2/3,1/3)$--and the most stable configuration follows a close-packing sequence, with the iodine atom closest to In$_2$Se$_3$ sitting on the hollow site of the nearby InSe sublayer while the neighboring Cu atom lies on top of the closest In (see insets in Fig.~\ref{fig:hetero}).

The energy bands for the In$_2$Se$_3$/CuI heterostructure in both polarization states, obtained with DFT-PBE with or without SOC, are shown in Fig.~\ref{fig:hetero}. When the polarization points from In$_2$Se$_3$ to CuI (denoted $\uparrow$,  Fig.~\ref{fig:hetero}a), the conduction band of In$_2$Se$_3$ lies above the valence band of CuI as anticipated from the relative alignment of the isolated layers, but the energy gap $E_{\rm g} = 0.36$~eV is much larger than the expected value $\Delta E^{\uparrow} = 0.02$~eV, i.e.\ $E_{\rm g} = \Delta E^{\uparrow}  + \delta^{\uparrow}$. The difference $\delta^{\uparrow}$ arises from several effects, but it can be mainly interpreted as a result of the modification of the wavefunctions close to the interface due to the repulsion from the other layer. The corresponding change in electronic density gives rise to an interface electric dipole that affects the relative alignment between valence and conduction bands and thus the energy gap. Moreover, we note that the electronic charge redistribution is from CuI to In$_2$Se$_3$, so that the overall out-of-plane polarization of the heterostructure is larger in magnitude than for isolated In$_2$Se$_3$.

When the polarization points from CuI to In$_2$Se$_3$ (denoted $\downarrow$), the interlayer distance is slightly smaller ($d_{\downarrow} = 3.04$~\AA) than in the previous case ($d_{\uparrow} = 3.10$~\AA). The corresponding band structure is reported in Fig.~\ref{fig:hetero}b. As expected, a band inversion is present, with the bottom of the conduction band associated with  In$_2$Se$_3$ lying lower in energy than the top of the valence band of CuI. Without SOC, the system is metallic with valence and conduction bands crossing at 6 symmetry-related Dirac points along the $\Gamma$-M directions. When SOC is included, an overall band gap of $52$~meV opens between valence and conduction bands. As a consequence of the band inversion, some valence band states in CuI get empty in favor of some conduction band states in In$_2$Se$_3$ that get occupied. This charge transfer from CuI to In$_2$Se$_3$ provides an additional contribution to the overall polarization of the heterostructure, which maintains the same direction but a reduced magnitude with respect to isolated In$_2$Se$_3$. The charge transfer also affects the band inversion  $E_{\rm i}$ at $\Gamma$, whose  value $E_{\rm i}=0.28$~eV differs from the expectation based on isolated monolayers $|\Delta E^{\downarrow}|= 1.1$~eV, i.e.\ $E_{\rm i} = |\Delta E^{\downarrow} + \delta^{\downarrow}|$ with $\delta^{\downarrow}>0$.

We have thus obtained that the magnitude of the vertical electric dipole in the two polarization states is not the same but $|P^{\uparrow}| > |P^{\downarrow}|$. As a consequence, we expect the hysteresis loop for the heterostructure to be asymmetric, as schematically depicted in Fig.~\ref{fig:scheme}. This asymmetry is reflected also in the relative stability between the two polarization states, for which we find the $\downarrow$ state slightly more stable than the $\uparrow$ state by $\sim 20$~meV.

\subsection*{Topological properties}

\begin{figure}
    \includegraphics[width=0.49\textwidth]{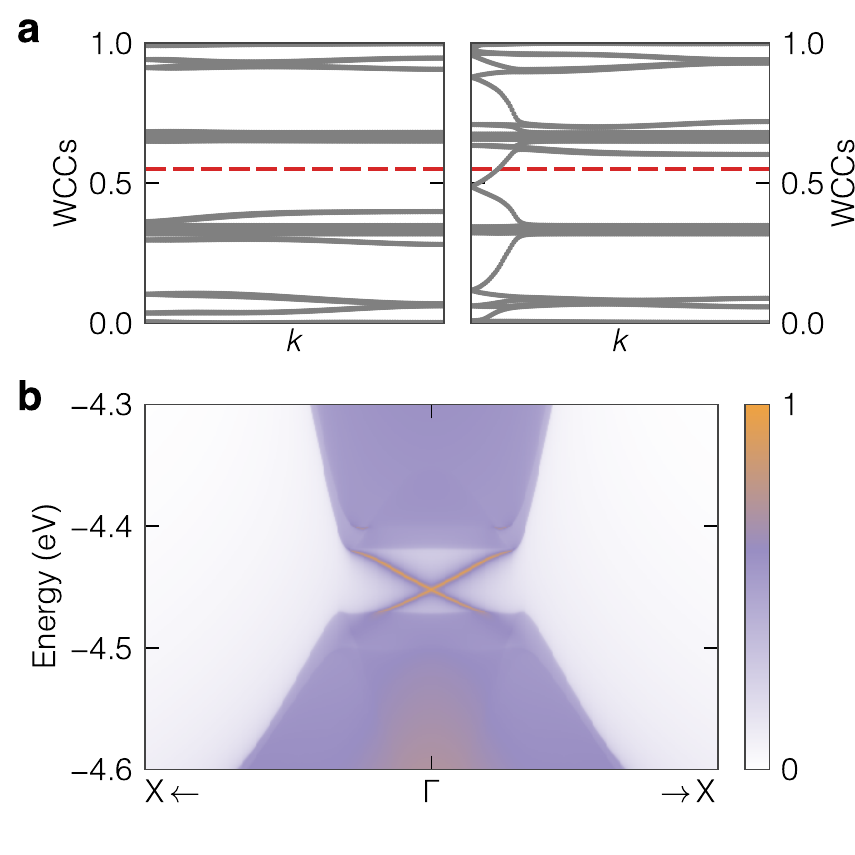}
    \titledcaption{Topological classification of the heterostructure}{\a\ Evolution of the hybrid Wannier charge centers~(WCCs) for the In$_2$Se$_3$/CuI heterostructure in the $\uparrow$ (left) and $\downarrow$ (right) polarization states. Any horizontal line such as, e.g., the red dashed line crosses an even (odd) number of times the WCCs in the former (latter) case. We thus have that when the polarization points from  In$_2$Se$_3$ to CuI ($\uparrow$, left) the system is trivial, while when it points from CuI to In$_2$Se$_3$ ($\downarrow$, right) the heterostructure is a quantum spin Hall insulator~(QSHI). \b\ Edge spectral density for a semi-infinite ribbon of In$_2$Se$_3$/CuI in the QSHIs phase, where a pair of helical edge states crosses the bulk band gap.}
    \label{fig:topo}
\end{figure}

The different band structure for the two polarization directions, with the presence of a band inversion in only one of them, suggests that the topological state of the heterostructure depends on polarization. To verify this expectation, in Fig.~\ref{fig:topo}a we show the computed evolution of the hybrid Wannier charge centers in the two cases, which allows to assess the $\mathbb{Z}_2$ topological invariant $\nu$ by counting the number of times $N$ any horizontal line crosses them as $\nu = (-1)^N$ \cite{Soluyanov_PRB_2011,z2pack_gresch_17}. When the polarization points  from In$_2$Se$_3$ to CuI ($\uparrow$), we have an even number of crossings, so that the invariant is $\nu_{\uparrow}=0$ and the material is trivial. On the contrary, when the polarization points in the opposite direction ($\downarrow$), we find an odd number of crossing, so that $\nu_{\downarrow}=1$ and the heterostructure is a topological insulator. We thus have that the heterostructure behaves as a ferroelectric quantum spin Hall insulator, where the polarization direction dictates the topological phase of the system, which can thus be manipulated in a non-volatile fashion by using an external electric field.

As a consequence of the non-trivial topology in the $\downarrow$-polarization state, we expect the presence of helical edge states that cross the bulk gap. In Fig.~\ref{fig:topo}b we show the  edge spectral density for a zigzag edge of the In$_2$Se$_3$/CuI heterostructure computed using a recursive Green's function approach~\cite{sancho_highly_1985} as implemented in WannierTools~\cite{WT_2018} (see Methods). Helical states inside the bulk gap are indeed clearly visible and disappear when considering the opposite (i.e.\ $\uparrow$) polarization (not shown). We notice that, since In$_2$Se$_3$ supports also a finite in-plane component of polarization, additional trivial edge states might appear in both $\uparrow$ and $\downarrow$ states depending on the edge  orientation and termination. While in Fig.~\ref{fig:topo}b the zigzag edge termination has been chosen to avoid such trivial edge states, they might appear for other zigzag terminations, while in case of armchair edges no trivial edge states are expected, suggesting that this orientation should be preferential for experimental investigations.

\subsection*{Role of vdW and SOC}

\begin{figure}
    \includegraphics[width=1\linewidth]{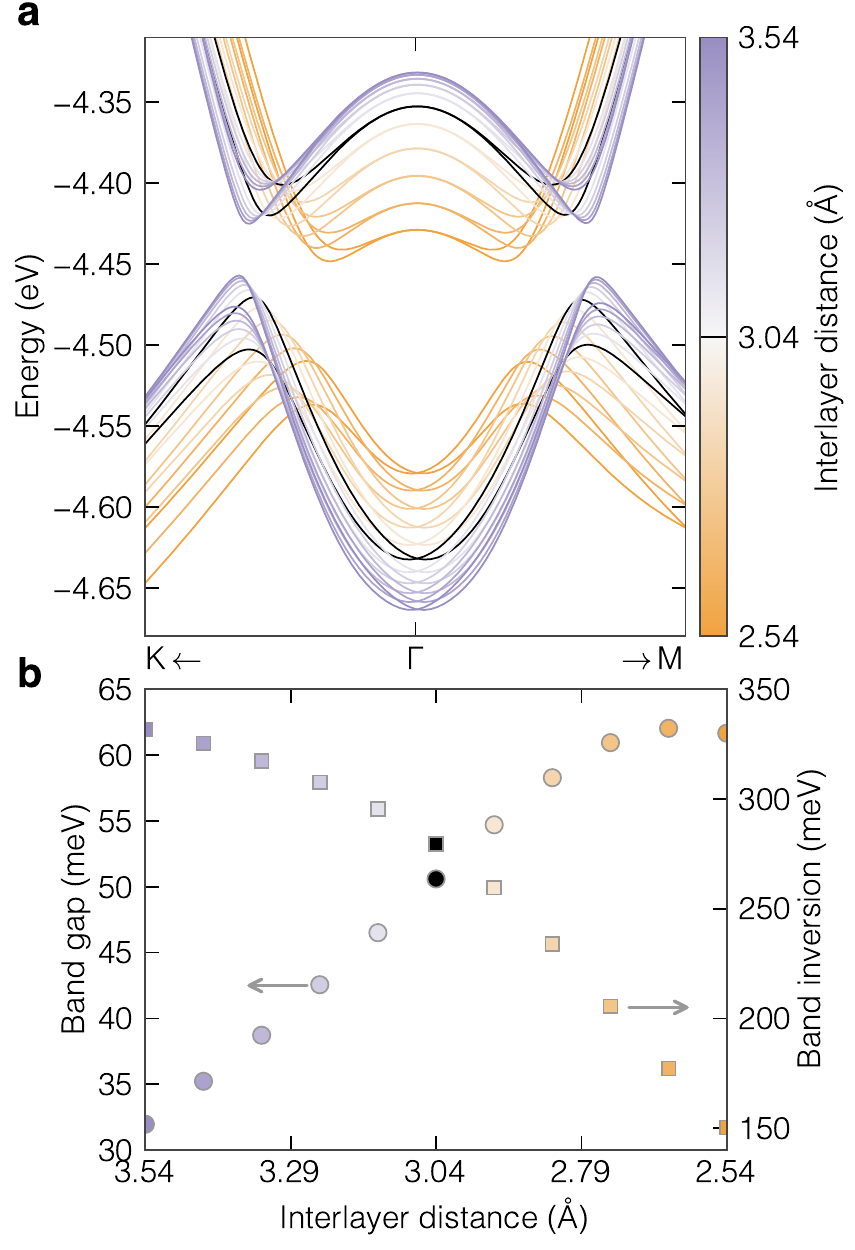}
    \titledcaption{Band structure evolution with interlayer distance}{\a\ Energy bands of the In$_2$Se$_3$/CuI heterostructure around $\Gamma$ computed at the PBE-DFT level including spin-orbit coupling at different interlayer distances (color codes) around the equilibrium value. \b\ Energy gap (circles, left axis) and band inversion at $\Gamma$ (squares, right axis) between valence and conduction bands as a function of interlayer distance. In \a\ and \b\ results for the equilibrium interlayer distance are reported in black. Upon compression, the band inversion decreases but the interlayer coupling strongly increases, so that the overall gap is enhanced.}
    \label{fig:displ}
\end{figure}

The robustness of a topological phase is typically measured by two quantities: the size of the energy band inversion and the magnitude of the band gap appearing at the crossings between the inverted bands\cite{marrazzo_relative_2019,custodialwte2_prb_2011,bernevig_topological_2013}. 
Notably, in ferroelectric heterostructures the strength of the band inversion can be made arbitrarily large by a suitable choice of materials, as it is dictated by the band alignment between them, and it is only limited by the potential drop across the ferroelectric layer through the requirement that the heterostructure is trivial for the opposite polarization.  Remarkably, we have seen that even the topological gap can be quite large ($\sim 50$~meV in the present case). We now want to show that such large band gaps are a general feature to be expected in these heterostructures as they are driven by a subtle interplay between the interlayer vdW hybridization and the intralayer SOC.  On one side, if at least one of the layers has a large SOC, a sizable band gap can appear despite the weak vdW nature of the interlayer coupling. On the other, if the materials involved in the heterostructure allow for a sufficiently small distance between the layers, the interlayer hybridization and thus the topological gap are enhanced. 

We start investigating the effect of interlayer coupling by first studying the evolution of the band structure around $\Gamma$ as a function of the interlayer distance around its equilibrium value, within a range of $\pm 0.5$~\AA. As reported in Fig.~\ref{fig:displ}a, the variation in interlayer separation gives rise to an almost rigid shift of the energy bands, leading to a reduction in the band inversion with decreasing interlayer distance as a result of a larger interface dipole upon compression. At the same time, the band gap opens closer to the $\Gamma$ point and increases in magnitude. A more quantitative analysis reported in Fig.~\ref{fig:displ}b shows that, when  the interlayer distance is reduced--and thus interlayer coupling is enhanced, the band gap increases steadily from $32$ up to $62$ meV, while the band inversion decreases from 340 to 150 meV. These large effects in response to a moderate change in interlayer distance suggest that interlayer coupling plays a crucial role in determining the band gap together with SOC and need further investigation. 

\begin{figure}
    \includegraphics[width=\linewidth]{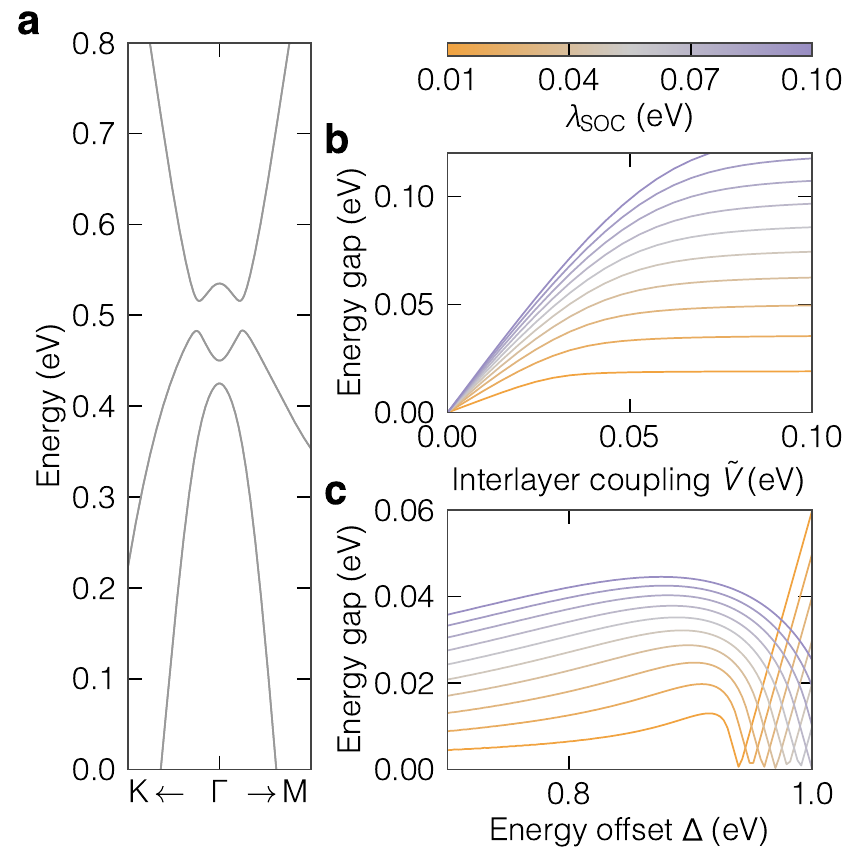}
    \titledcaption{Tight-binding model}{\textbf{a} Band structure around $\Gamma$ for the Slater-Koster $s-p_xp_y$ tight-binding model discussed in the text, where a band inversion between the second and third band occurs at the Fermi level and is gapped by spin-orbit coupling (SOC). \textbf{b} Band gap as a function of the interlayer coupling $\tilde{V}$ for various SOC strengths ($\lambda_{\rm SOC}$, color-coded lines): the gap increases linearly with $\tilde{V}$ until it saturates to a value of the order of $\lambda_{\rm SOC}$. \textbf{c} Band gap as a function of the energy offset $\Delta$ between orbitals on different layers for various values of $\lambda_{\rm SOC}$(same color coding as panel \textbf{b}): in the topological phase, the gap moderately increases with the energy offset while the band inversion is reduced, until it suddenly drops to zero when the band inversion vanishes. The gap then reopens and increases linearly with $\Delta$ in the trivial insulating phase.}
    \label{FigModel}
\end{figure}

To disclose the origin of these phenomena and to assess their general validity, we introduce a Slater-Koster~\cite{SlaterKoster_prb_1954} tight-binding (TB) model that qualitatively reproduces the band structure around the Fermi level (see Supplementary Note 3). The model is composed of an $s$-like orbital localized on the In$_2$Se$_3$ layer and of $p_x,p_y$-orbitals localized on CuI, so that the centers of all orbitals are vertically aligned and their position differ only by the $z$-coordinate of the two layers (see Supplementary Fig.~3). Beyond the intralayer nearest-neighbour hopping terms that set the effective mass of the energy bands close to $\Gamma$, the model includes  the energy offset $\Delta$ between the orbitals in the two layers and an interlayer nearest-neighbour hopping $\tilde V$ between $s$ and $p_x,p_y$-orbitals that is responsible for the interlayer hybridization. SOC is included only on CuI through an on-site term with strength $\lambda_{\rm SOC}$. Fig.~\ref{FigModel} shows the model band structure around $\Gamma$ in the QSHI phase, with realistic parameters (see Supplementary Table 2) that reproduce qualitatively the first-principles results in Fig.~\ref{fig:hetero}b. 

Notably, although $\lambda_{\rm SOC}$ is totally localized on the $p_x,p_y$-orbitals, it is still able to open a topological band gap between bands belonging to well separated layers. In fact, we now want to show that the band gap opening is due to an on-site SOC--localized on a single layer--that is \emph{mediated} by the interlayer coupling $\tilde{V}$. We thus compute the band gap as a function of the interlayer interaction, as shown in Fig.~\ref{FigModel}b. In the limit of non-interacting layers, i.e. $\tilde{V}\rightarrow0$,  there is no band gap opening, independently of the SOC strength, suggesting that indeed the degeneracy at the crossing point can be lifted only if there is some hybridization between the orbitals sitting on the two layers. In the regime where $\tilde{V}\lesssim \lambda_{\rm SOC}$, as it is the case for In$_2$Se$_3$/CuI at equilibrium, the band gap depends linearly on the interlayer coupling, which means that increasing the interaction between the layers (e.g. by reducing the interlayer distance) greatly improves the band gap in the QSHI phase. If $\tilde{V}\gtrsim\lambda_{\rm SOC}$, then the band gap still increases with the interlayer distance but saturates at a value proportional to $\lambda_{\rm SOC}$ (the exact prefactor depends on the value of the other TB parameters). Remarkably, the interlayer hopping does not suppress the effect of SOC but it rather allows to achieve band gaps comparable (if not higher) to the SOC strength $\lambda_{\rm SOC}$. This effect is similar to the orbital filtering obtained in honeycomb lattices with $p_x,p_y$-orbitals~\cite{orbitfilt_prb_2014}, such as Bi on SiC~\cite{reis_science_2017}, although the mechanism there is different and, in particular, the topological band gaps discussed in Ref.~\onlinecite{orbitfilt_prb_2014} are equal to the SOC strength only at the special point K~\cite{orbitfilt_prb_2014}. Here, instead, the band gap is of the same order of magnitude of $\lambda_{\rm SOC}$, but it appears in a low-symmetry point around $\Gamma$.
These results closely match the first-principles simulations shown in Fig.~\ref{fig:hetero}, providing the following picture: as the interlayer distance is reduced, the interlayer hopping increases correspondingly such that the band gap of the QSHI phase increases, first linearly and then saturating to a value of the order of $\lambda_{\rm SOC}$.

Actually, reducing the interlayer distance might affect the band gap also through the orbital energy offset $\Delta$, although this effect is much weaker as it is shown in Fig.~\ref{FigModel}c. If $\Delta$ decreases, the band inversion becomes stronger and the crossing point between the conduction and valence bands moves farther from the $\Gamma$ point. The effect of SOC becomes smaller as the crossing point moves away from $\Gamma$, resulting in a decreasing band gap. Viceversa, increasing the energy offset leads to a larger band gap and smaller band inversion, as long as the system remains a QSHI: if $\Delta$ becomes too large then the band inversion disappears, the gap quickly drops to zero before increasing again with the offset as the system has entered into the trivial insulating phase.

\begin{figure*}
    \includegraphics[width=\linewidth]{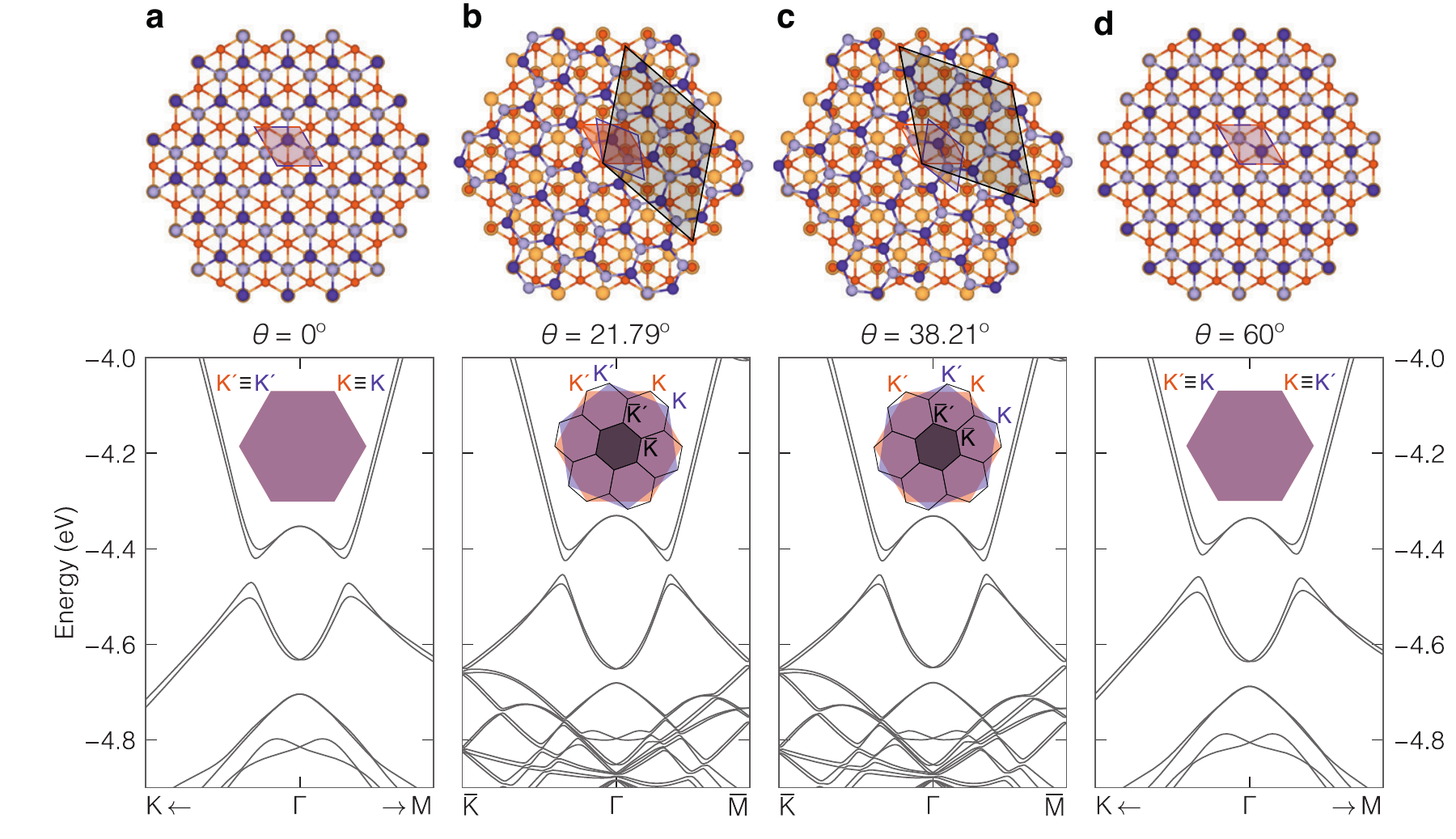}
    \titledcaption{Effect of relative orientation between the layers}{Band structure of the In$_2$Se$_3$/CuI heterostructure close to the Brillouin zone center for different values of the relative rotation angle $\theta$ between the layers: $0^\circ$ (\a), $21.79^\circ$ (\b), $38.21^\circ$ (\c), and $60^\circ$ (\d). For each case, a top view of the heterostructure is reported, where the primitive unit cells of In$_2$Se$_3$, CuI, and the heterostructure are shown as orange, violet, and black shaded areas, respectively. For $\theta=0$ and $60^\circ$ the heterostructure displays the same translational symmetry of the two separate layers and can thus be accommodated within a single primitive cell. On the contrary, for $\theta = 21.79^\circ$  and $38.21^\circ$ a $\sqrt{7}\times\sqrt{7}$ supercell is needed. In all cases, a finite energy gap is present between the valence and conduction bands and the system is a quantum spin Hall insulator, showing that the relative alignment between the layers has no effect on the topological classification of the heterostructure.  Insets show the hexagonal Brillouin zone of In$_2$Se$_3$ (orange), CuI (violet), and the heterostructure (black), with the inequivalent corners, K and K$'$, highlighted.
    \label{fig:twist} }
\end{figure*}

\subsection*{Role of relative rotation angle}

Up to now we have considered a primitive unit cell and perfectly aligned lattices for the two layers, thanks to the lattice matching between CuI and In$_2$Se$_3$. We now want to argue that lattice matching and crystalline alignment are not necessary, and that the topological state is preserved even considering non-primitive unit cells arising, e.g., from a relative rotation between the layers. We thus consider four different twist angles $\theta$ between In$_2$Se$_3$  and  CuI: $0^\circ$, $21.79^\circ$, $38.21^\circ$ and $60^\circ$. While for $\theta= 0^\circ$ and $60^\circ$ the heterostructure exhibits the same translational symmetry of the two layers and can be accommodated within a  single primitive cell, for $\theta =21.79^\circ$ and $38.21^\circ$ a $\sqrt{7}\times\sqrt{7}$  supercell with 63 atoms is necessary to account for the relative orientation between the layers (see Methods for more details).

In Fig.~\ref{fig:twist} we report the band structures calculated by PBE-DFT first-principles simulations with SOC at the four different twist angles. Remarkably, the heterostructure remains a QSHI with a finite indirect band gap at all twist angles, showing that changes in the relative orientation between the layers do not undermine the topological phase. This is due to the fact that the band inversion occurs at the BZ center $\Gamma$, and so it is relatively insensitive to the twist angle. In particular, the SOC-induced gap is only weakly affected, with a value at $\theta = 21.79^\circ$ and $38.21^\circ$ of 29 meV,  close to the 52 meV that is obtained at $0^\circ$ and $60^\circ$. Correspondingly, also the band inversion is almost unaffected  by the twist angle, with a marginal increase for  $\theta = 21.79^\circ$ and $38.21^\circ$ with respected to perfect alignment. These very weak effects on the band structure can be accounted for by a slight increase in interlayer distance arising from the twist angle that does not allow an ideal close-packing configuration. In agreement with Fig.~\ref{fig:displ}b, an increase in separation between the layers leads to a slight increase in band inversion and to a decrease in band gap, also due to a reduction in the effective interlayer coupling associated with the misalignment.

\subsection*{Role of layer thickness}

\begin{figure}
    \includegraphics[width=1\linewidth]{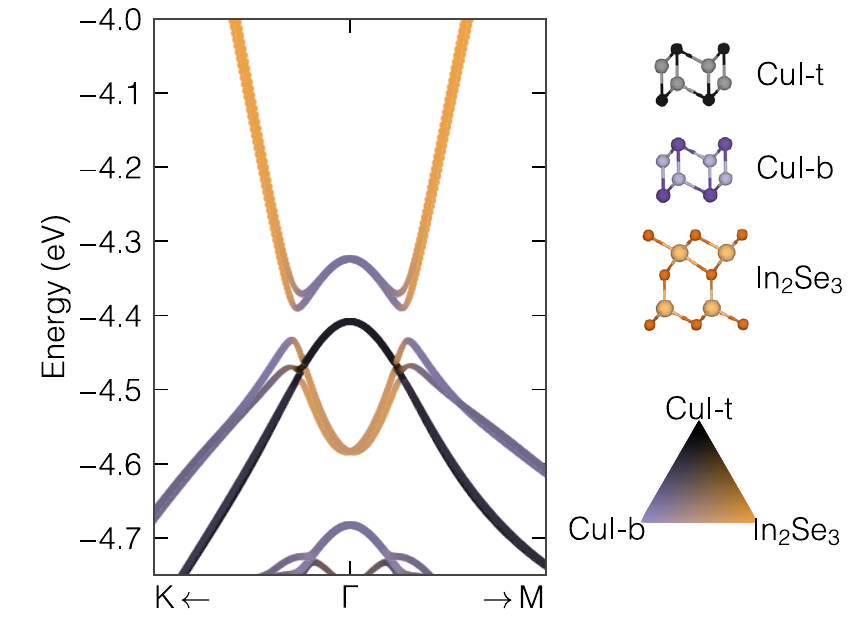}
    \titledcaption{Band structure of In$_2$Se$_3$-bilayer CuI}{Energy bands around the Brillouin zone center for a heterostructure comprising monolayer In$_2$Se$_3$ and bilayer CuI (lateral view on the right). The color coding represents the layer contribution to electronic states, with orange denoting In$_2$Se$_3$, violet the bottom CuI closer to the ferroelectric layer, and black the upper CuI. A band inversion between the conduction band of In$_2$Se$_3$ and the valence bands of CuI is still present, with a strong hybridization between In$_2$Se$_3$ and the bottom CuI giving rise to a topological gap, while the valence band of the top CuI layer (black) is largely unaffected.}
    \label{fig:bilayer}
\end{figure}

Here we comment on the possibility of observing this phenomenon even when the heterostructure is composed of materials with more than a single layer. First, we expect that  tunneling (i.e.\ hopping) between the layers in each material will lead to the splitting of the valence and conduction bands into subbands, thus affecting the band alignment. Another important effect arises from  the relative orientation of the out-of-plane polarization, which can be either parallel or antiparallel, when multiple layers of the ferroelectric material are stacked together. If it is parallel, the potential drop associated with each layer will add up and would potentially lead to a ``polar catastrophe''~\cite{nakagawa_why_2006} with an increasing layer thickness, which is prevented by an electronic reconstruction and the appearance of metallic states on the top and bottom surface of the material. According to previous simulations~\cite{ding_prediction_2017}, this should occur already in bilayer In$_2$Se$_3$ and would hinder the observation of the predicted effect as the metallic surface states screen the potential drop arising from bulk polarization. Nonetheless, a parallel configuration seems  experimentally unlikely~\cite{cui_intercorrelated_2018}, in favor of an antiparallel configuration. In this case, for an even number of layers the polarization is perfectly compensated and there would be again no potential drop. Still, for an odd number of antiparallel layers, the polarization is necessarily uncompensated, with a potential drop essentially equivalent to the one of a monolayer. We thus expect a band inversion driven by the potential drop to be in principle still observable when instead of a single layer we have an odd-layer (anti)ferroelectric. 
 
Even in this case, there might be subtle effects associated with the thickness of the semiconducting material. Provided that the subband dispersion is not too large, we still expect the system to be a trivial insulator in one polarization state irrespective of the number of layers. With the opposite polarization, a band inversion might still occur between a band of the ferroelectric material and possibly multiple (sub)bands of the semiconductor. The resulting charge transfer is likely to be localized on the layers closest to the interface as a result of self-consistent electrostatic screening effects. This pronounced inequivalence between the interface layers and the outer ones, which are farther from the interface, is reflected in a strong localization of subbands on the interface layers, which can hybridize through the vdW gap between the materials. The combined effect of such vdW coupling and SOC opens a gap between these interface subbands, while leaving essentially unaffected the other subbands that have a marginal contribution from the interface. We thus expect even in thicker systems to be able to observe the same physical phenomena described above, with interface layers playing the role of the monolayers. 

We have verified this picture for the specific case of In$_2$Se$_3$/CuI by performing first-principles simulations for a heterostructure made of two layers of CuI and one layer of In$_2$Se$_3$. Fig.~\ref{fig:bilayer} shows that a band inversion between the conduction band of In$_2$Se$_3$ and the valence bands of CuI is still present. As expected, bands have a strong layer localization and a significant vdW hybridization occurs only between the interface bands opening up a gap in the spectrum, while the valence band associated with the top CuI layer (black) is largely unaffected and remains completely filled. Thus, although at the $\Gamma$ point the conduction band of In$_2$Se$_3$ is below both bands of CuI, the charge transfer and band hybridization happens only between the interface layers. 
Given the weak vdW coupling between CuI layers, we expect the same to be true also for thicker CuI,  suggesting that it should be possible to realize a ferroelectric QSHI even by deposing monolayer In$_2$Se$_3$ on the cleaved surface of a bulk CuI sample, which is experimentally even more feasible than the heterostructure made of monolayers. We remark that the band hybridization occurs only for the CuI layer exposed at the interface with In$_2$Se$_3$, hence only the interface composed by one CuI layer and one layer of In$_2$Se$_3$ will be a topological insulator, while the rest of CuI remains a trivial semiconductor.

\section*{Discussion}

In this work, we have shown how robust ferroelectric quantum spin Hall states can appear in van-der-Waals heterobilayers that either occur spontaneously (In$_2$ZnS$_4$) or by design (In$_2$Se$_3$/CuI), where the topological phase of the system can be controlled reversibly and in a non-volatile way through the ferroelectric polarization direction. Remarkably, the topological gap arises from a combination of intralayer spin-orbit coupling and interlayer hybridization, leading to significantly large values despite the weak nature of van der Waals interactions. Even more compelling, we have demonstrated that, when the band extrema in the two materials composing the heterostructure lie at the Brillouin zone center, the effect is resilient to the relative orientation between the layers and does not require lattice matching. In addition, we verified that the band inversion persists even if a CuI bilayer is considered, suggesting that a single layer of In$_2$Se$_3$ deposited on the surface of a thick CuI sample is sufficient to obtain a 2D ferroelectric quantum spin Hall insulator. The proposed mechanism is thus very general and requires only a proper band alignment between the conduction and valences states of two monolayers, a ferroelectric and a semiconductor.  Considering  the extensive  portfolio of 2D materials potentially available~\cite{mounet_two-dimensional_2018,haastrup_c2db_2018,gjerding_recent_2021}, there is a combinatorially large number of heterostructures to be explored in experiments, possibly leading to even more robust topological phases and more complex interplays between ferroelectricity and topology.

\section*{METHODS}
\subsection*{First-principles simulations}
DFT calculations are performed with the Quantum ESPRESSO distribution~\cite{giannozzi_quantum_2009,Giannozzi2017}, using the PBE functional~\cite{perdew_pbe_96} and the PseudoDojo~\cite{ONCVPSP,dojo_paper_18} pseudopotential library.  The wavefunction and charge density energy cutoffs used to simulate the In$_2$Se$_3$/CuI heterostructure are set to $100$ Ry and $400$ Ry, respectively. The Brillouin zone is sampled using a regular $\Gamma$-centered Monkhorst-Pack grid with $12\times12\times1$ k-points, with a small cold smearing of $7.3\times10^{-3}$ Ry for the topological heterostructures. A Coulomb cutoff~\cite{Rozzi2006,Sohier2017} is used to avoid spurious interactions between periodic replicas and thus simulate the correct boundary conditions for 2D system. Structural relaxations are performed without spin-orbit coupling using the revised Vydrov-Van Voorhis (rVV10) non-local van-der-Waals functional~\cite{vydrov_vv10_09,sabatini_rvv10_13}. Band structures are then computed on top of the relaxed structure including spin-orbit coupling through fully-relativistic pseudopotentials. 
Maximally-localized Wannier functions are obtained using WANNIER90~\cite{mostofi_updated_2014,Pizzi_2020}, tight-binding models are created with PythTB \footnote{https://www.physics.rutgers.edu/pythtb/index.html}, and the edge spectral density is calculated with WannierTools~\cite{WT_2018}.
Topological invariants are computed using Z2Pack~\cite{soluyanov_z2pack_11,z2pack_gresch_17} and WannierTools~\cite{WT_2018}.

Hybrid-functional calculations have been performed using the Heyd-Scuseria-Ernzerhof (HSE) functional~\cite{HSE} as implemented in Quantum ESPRESSO~\cite{Giannozzi2017} with the acceleration provided by the Adaptively Compressed Exchange Operator~\cite{lin_ace_2016}. A cutoff of 100 Ry (equal to the wavefunction cutoff) for the Fock operator has been sufficient to converge self-consistent band energies, with a q-grid of $12\times12\times1$ ($6\times6\times1$) for topological (trivial) systems. Results on the irreducible Brillouin zone are expanded to the full zone using the \textsf{open\_grid.x} code in Quantum ESPRESSO~\cite{Giannozzi2017}, and then band structures are  interpolated using a Wannier representation. 

G$_0$W$_0$ calculations are performed using the Yambo~\cite{yambo_2019} code, on top of DFT-PBE calculations with the Quantum ESPRESSO distribution~\cite{giannozzi_quantum_2009,Giannozzi2017}. We use fully relativistic ONCV~\cite{ONCVPSP} pseudopotentials from the SG15 library~\cite{sg15_2015}. The self-energy is constructed using a $36\times36\times1$ k-point grid. In the G$_0$W$_0$ calculations we adopt the random integration method, the 2D Coulomb cutoff, the Bruneval-Gonze terminator~\cite{bg_prb_2008} for the Green's function and the Godby-Needs~\cite{godbyneeds_prl_1989} plasmon pole approximation for the frequency dependence of the self-energy. SOC is included self-consistently at the DFT level, using spin-orbitals, and fully taken into account at the G$_0$W$_0$ level using a spinorial Green's function.

\subsection*{Band alignment}

In 3D materials, band energies are computed with respect to a material-dependent reference value, thus making direct comparison between band energies in different materials ill-defined and the evaluation of band offsets rather intricate. The situation is simplified in 2D materials, where a well-defined reference energy can be obtained by considering the constant limiting  value of the total electrostatic potential reached far away from the material, i.e. the so-called vacuum energy. By shifting the band energies so that this reference vacuum energy is the same for both materials, we obtain the correct band alignment between different materials. The procedure is further simplified when calculations are performed using a cutoff to truncate Coulomb interactions along the vertical direction~\cite{Rozzi2006,Sohier2017}, orthogonal to the layers, which allows one to mimic the correct open-boundary conditions of the 2D system even though the calculations are performed using a plane-wave basis set, and thus with periodic boundary conditions in all directions. As a result, band energies are referred to a well-defined value, with a reference zero set at the vacuum level of a neutral non-polar system. It is thus very easy to compare band structures of different non-polar materials on an absolute scale, and the relative band alignment can be obtained by directly comparing the bare band energies. When a material (or both) has a finite vertical dipole, the total electrostatic potential in the vacuum does not go to the reference zero, but to two opposite values on the two sides of the material ($\pm\Delta\phi$ for In$_2$Se$_3$ in the main text). The correct band offset between two materials can thus be obtained from the bare band energies by correcting for the electrostatic offset needed to re-align the vacuum energies in the region between the materials. The alignment thus obtained corresponds to having the materials sufficiently far apart to avoid charge transfer or any other source of charge redistribution that can further affect the band offset when the materials are instead close enough.

\subsection*{Supercell creation}
Supercells for twisted heterostructures are created by considering larger, non-primitive unit cells for the two layers defined by  a first lattice  vector  $\mathbf{A}^{\rm In_2Se_3}_1 = n_{1}\mathbf{a}_1 + n_{2} \mathbf{a}_2$ and  $\mathbf{A}^{\rm CuI}_1 = m_{1}\mathbf{a}_1 + m_{2} \mathbf{a}_2$, where $\mathbf{a}_{1,2}$ are the common primitive lattice vectors of In$_2$Se$_3$ and CuI,  while the other lattice vector is obtained by a $120^\circ$ rotation. The volume of these unit cells is increased by a factor $n_1^2 + n_2^2 - n_1 n_2 =  m_1^2 + m_2^2 - m_1 m_2$. The CuI cell is then rotated to align $\mathbf{A}^{\rm CuI}_1$ onto  $\mathbf{A}^{\rm In_2Se_3}_1$ and have a common Bravais lattice for the supercell. The simple primitive case is recovered for $n_1=m_1=1$ and $n_2=m_2=0$, while the $\theta = 60^\circ$ case is obtained for $(n_1,n_2)=(1,0)$ and $(m_1,m_2)=(1,1)$. The other rotation angles considered in the main text correspond to a 7-fold supercell with $(n_1,n_2)=(2,-1)$ and $(m_1,m_2)=(3,1)$ for $\theta = 38.21^\circ$, while $(n_1,n_2)=(1,-2)$ and $(m_1,m_2)=(2,-1)$ for $\theta =21.79^\circ$.

\section*{DATA AVAILABILITY}
The data that support the findings of this study are available from the corresponding author upon reasonable request.

\section*{CODE AVAILABILITY}
The electronic structure codes used in this work are all open source and available online at their corresponding website. Input files, tight-binding models and other relevant scripts are available from the corresponding author upon reasonable request.

\bibliographystyle{mynaturemag}
\bibliography{Ferrotopo}

\section*{Acknowledgements}
The authors would like to thank Nicola Marzari for useful discussions.  We acknowledge support during the initial phase of the project from the NCCR MARVEL (A.M. and M.G.) and the Ambizione program (M.G.), both funded by the Swiss National Science Foundation. M.G. acknowledges support from the Italian Ministry for University and Research through the Levi-Montalcini program.  Simulation time was  awarded by PRACE (project id. 2016163963), ISCRA and a CINECA-UniTS agreement on MARCONI100 at CINECA, Italy.

\section*{Author contributions}
M.G. originally conceived the project based on a materials discovery by A.M.. A.M. and M.G. together further developed the project, performed simulations and modeling, and wrote the manuscript.

\section*{Competing Interests}
The authors declare no competing interests.
\clearpage

%\end{document}
\onecolumngrid
  
\section*{\LARGE\bfseries Supplementary Information}

\renewcommand{\figurename}{Supplementary Fig.}
\renewcommand{\tablename}{Supplementary Tab.}
\renewcommand\thefigure{\arabic{figure}}
\renewcommand\thetable{\arabic{table}}

\renewcommand\thesubsection{Supplementary Note \arabic{subsection}}

\setcounter{equation}{0}
\setcounter{figure}{0}
\setcounter{table}{0}
\setcounter{subsection}{0}
\newcommand\kp{{\bm k}_\parallel}

\subsection{Candidate materials}

As discussed in the main text, in order to realize a van der Waals (vdW) heterostructure that works as a ferroelectric Quantum Spin Hall Insulator (QSHI), the two isolated monolayers need to satisfy some stringent conditions: i) the band alignment between the valence band maximum $E_{\rm v}$ in one material and the conduction band bottom $E_{\rm c}$ in the other needs to be such that $|E_{\rm c}-E_{\rm v}| < \Delta\phi/2$, where $\Delta \phi$ is the magnitude of the  vacuum-level  discontinuity across the ferroelectric layer; ii) there must be a robust spin-orbit coupling (SOC) in one of the two layers, typically due to the presence of relatively heavy chemical elements. In order to make the effect resilient against twist rotations and/or lattice mismatch (both in terms of lattice parameters and symmetry), it is convenient that iii) both the conduction band bottom of one material and the valence band maximum of the other lie at the Brillouin zone (BZ) center, usually denoted as $\Gamma$ point. 

Taking for definiteness the ferroelectric layer to be In$_2$Se$_3$, for which $E_{\rm c}$ indeed lies at the $\Gamma$ point, we have screened large databases of 2D materials\cite{mounet_two-dimensional_2018,haastrup_c2db_2018,gjerding_recent_2021} for an optimal candidate with the top of the valence band satifying the conditions i) and iii) above. While these conditions might seem very strict, in reality there are many candidates that can satisfy them, as shown in the Supplementary Fig.~\ref{fig:align}, where the band alignment for several candidate two-dimensional (2D) materials is reported. Indeed in Supplementary Fig.~\ref{fig:align}, the filled bars mark the range of the valence bands for several 2D materials that have been selected among the semiconducting ones with the top of the valence bands $E_{\rm v}$ at the BZ center, corresponding to condition iii). In order to fulfill condition i), $E_{\rm v}$ needs to lie in the interval between $E_{\rm c}-\Delta\phi/2$ and $E_{\rm c}+\Delta\phi/2$, which is marked with solid orange lines and contains several prospective candidates. We note in passing that if one considers a different choice for the ferroelectric other than In$_2$Se$_3$, the number of candidates increases considerably more, such that the probability of finding even more promising heterostructures becomes very high.

As discussed in the main text, condition i) suffers from some drawbacks because it is based on the alignment between isolated monolayers and does not include interfacial effects like interface dipoles. As a consequence, some materials falling inside the highlighted range between solid lines in Supplementary Fig.~\ref{fig:align} might actually be false positive, and similarly materials that are excluded might be false negative. A refined condition that takes into account approximately the effect of interface dipoles might be $E_{\rm c}  - \Delta\phi/2  + \delta^\downarrow \le E_{\rm v} \le E_{\rm c}  + \Delta\phi/2 + \delta^\uparrow$, where $\delta^{\uparrow,\downarrow}$ is the energy shift associated with the interface dipole in the two polarization states. Assuming for simplicity, $\delta^\uparrow=\delta^\downarrow=\delta = 0.25$~eV, the corresponding refined range of interest is emphasized with orange dashed lines in Supplementary Fig.~\ref{fig:align}.

\begin{figure*}[b]
    \includegraphics{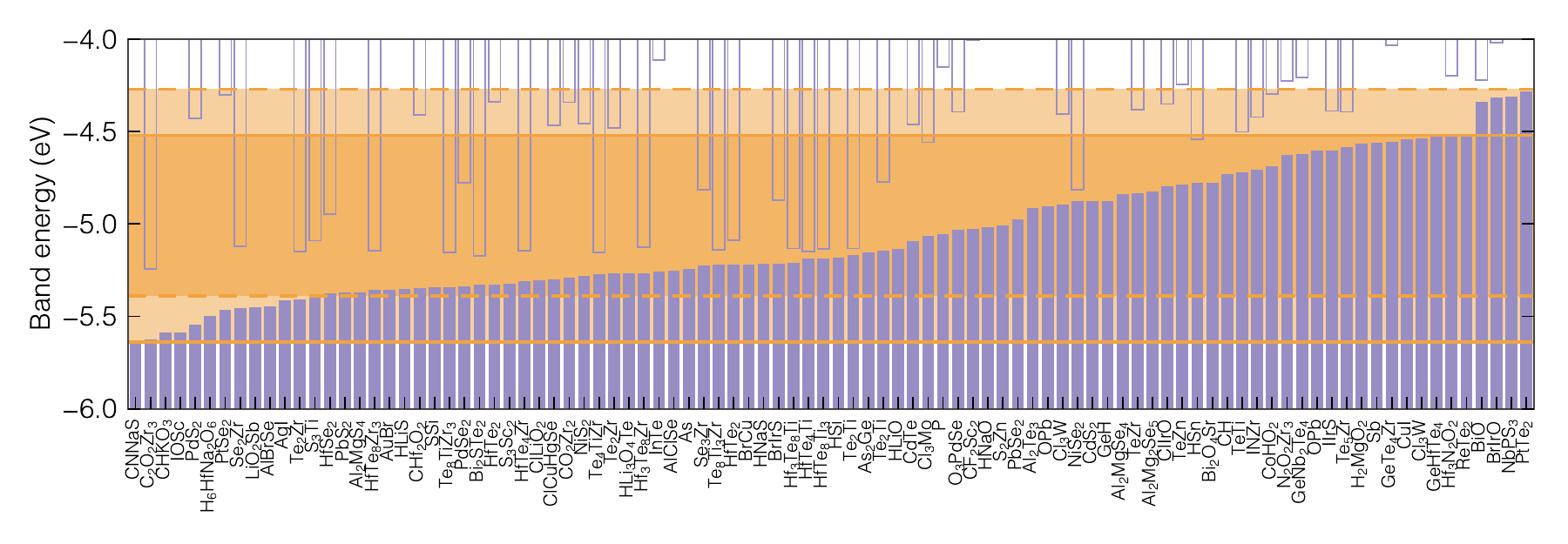}
    \caption{Band alignment of the valence bands (violet filled bars) for  several candidate 2D materials with respect to the bottom of the conduction band in the two polarization states $E_{\rm c}^{\uparrow,\downarrow} = E_{\rm c}  \pm \Delta\phi/2$  of In$_2$Se$_3$ (orange solid lines). For completeness, also the conduction band alignment for each 2D material is shown with violet empty bars. Candidate materials are selected from existing databases of 2D materials~\cite{mounet_two-dimensional_2018,haastrup_c2db_2018,gjerding_recent_2021} by considering semiconductors or insulators with energy gap of at least 0.1 eV with top of the valence band located at the Brillouin zone center, with an energy $E_{\rm v}$  having a promising alignment with the conduction band of In$_2$Se$_3$ to form a ferroelectric QSHI. This criterion corresponds to having $E_{\rm c}  - \Delta\phi/2 \le E_{\rm v} \le E_{\rm c}  + \Delta\phi/2$ (i.e. $E_{\rm v}$ between the orange solid lines), or by considering the effect of an interface dipole with $\delta^\uparrow=\delta^\downarrow=\delta = 0.25$~eV to having  $E_{\rm c}  - \Delta\phi/2  + \delta \le E_{\rm v} \le E_{\rm c}  + \Delta\phi/2 + \delta$ (i.e. $E_{\rm v}$ between the orange dashed lines).}
    \label{fig:align}
\end{figure*}

Among the materials falling in this refined range, which include HfTe$_2$~\cite{wang_tunable_2021}, As~\cite{zhang_heterobilayer_2021} and PtTe$_2$~\cite{phd_marrazzo_2019}, we further look for systems with sufficiently heavy elements in agreement with condition ii) and that are expected to be easily exfoliable from bulk parent materials~\cite{mounet_two-dimensional_2018}. To facilitate simulations, we also focus on materials that are approximately lattice matched with In$_2$Se$_3$, although this conditions is not needed for experimental realizations. Among these, in the main text we consider CuI to illustrate the effect and the possible realization of a ferroelectric QSHI. We stress that, although CuI is certainly interesting in view of the recent experimental progress in the growth of single layers~\cite{mustonen_towards_2021}, the specific choice is dictated entirely by illustrative purposes and that other materials could be equally relevant. We thus believe that there is an entire portfolio of prospective heterostructures for experimental investigations. In this respect, it is important to bear in mind that the current selection of potential candidates is based on density-functional-theory (DFT) calculations within routine approximations for the exchange-correlation functional. The accuracy of the calculated band alignment needs thus to be further tested with more sophisticated methods, as approximate DFT tends to underestimate band gaps and work functions. In \ref{app:beyond} we perform such analysis for CuI/In$_2$Se$_3$, with a partially positive assessment that indeed this heterostructure could give rise to a ferroelectric QSHI. Similar investigations could be performed also for other prospective systems and would very likely provide an ultimate candidate heterostructure. However, such investigation is computationally very demanding and beyond the illustrative scopes of the current study.

\subsection{Beyond DFT calculations}\label{app:beyond}

DFT, augmented with non-local vdW functionals and the inclusion of SOC, is capable of accurately predicting several properties relevant to these systems, such as the structural configuration of the heterostructure, the presence of charge transfer effects and the appearance of a topological phase. At the same time, it is well known that DFT cannot predict accurately spectral properties such as the band gap or the band alignment. In particular, DFT typically underestimates the band gap of insulators as well as their work function, for a 2D semiconductor the latter corresponds to minus the energy of the top of the valence band. 
Hence, although the main focus of this work is to discuss the general mechanism in a realistic setting more than proposing a specific material, we investigate how the first-principles prediction is affected by considering beyond-DFT methods such as hybrid functionals and many-body perturbation theory (MBPT) in the G$_0$W$_0$ approximation. A summary of the results in reported in Supplementary Table \ref{tab:beyond}.

\begin{table}[h]
\caption{\label{tab:beyond} Summary of the relevant absolute band energies and differences (in eV), comparing density functional theory at the PBE level with hybrid-functional simulations (specifically HSE) and  many-body-perturbation-theory results (within the \ G$_0$W$_0$ approximation). We include results for: the top of the valence band of isolated CuI ($E_{\rm v}$), the bottom of the conduction band of isolated In$_2$Se$_3$ ($E_{\rm c}$), the magnitude of the vacuum level difference across In$_2$Se$_3$ ($\Delta\phi$), the difference $E_{\rm c} - E_{\rm v}$, the (trivial/topological) energy gap of the CuI/In$_2$Se$_3$ heterostructure in  $\uparrow/\downarrow$ polarization state ($E_{\rm g}^{\uparrow/\downarrow}$), and the magnitude of the band inversion at the $\Gamma$ point in the topological phase with $\downarrow$ polarization ($E_{\rm inv}$).} 
\begin{tabularx}{0.8\linewidth}{>{\centering\arraybackslash}X >{\centering\arraybackslash}X >{\centering\arraybackslash}X >{\centering\arraybackslash}X >{\centering\arraybackslash}X >{\centering\arraybackslash}X >{\centering\arraybackslash}X >{\centering\arraybackslash}X}
\toprule
 & $E_{\rm v}$ & $E_{c}$ & $\Delta\phi$ & $ E_{\rm c} - E_{\rm v} $   & $E_{\rm g}^\uparrow$ & $E_{\rm g}^\downarrow$ & $E_{\rm inv}$ \\
 \midrule
 PBE & $-4.54\phantom{-}$ & $-5.08\phantom{-}$ & 1.12 & $-0.54\phantom{-}$ & 0.361 & 0.050 &  0.280\\
 HSE & $-5.25\phantom{-}$ & $-5.11\phantom{-}$ & 1.26 & $0.14$ & 1.046 &0.028 & 0.044\\
 G$_0$W$_0$ & $-6.41\phantom{-}$ & $-4.99\phantom{-}$ & 1.12 & 1.42  & --  & -- & -- \\
\bottomrule
\end{tabularx}
\end{table}

\begin{figure*}
    \includegraphics{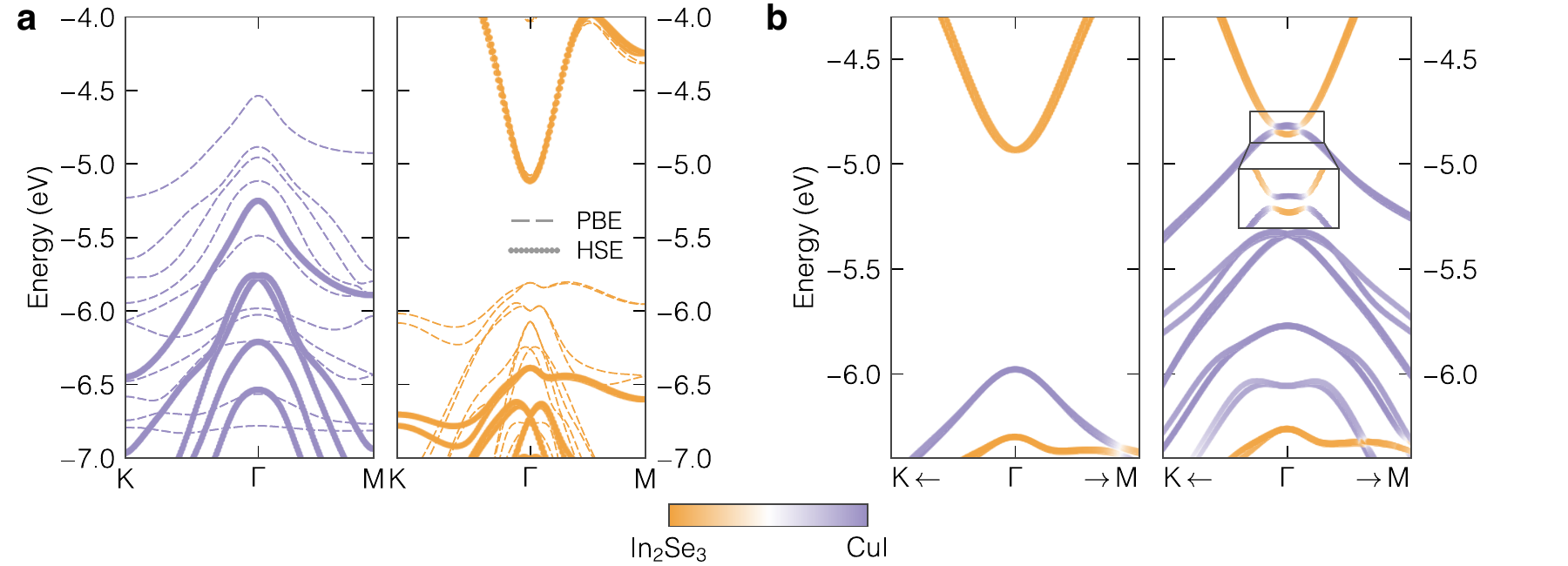}
    \caption{Calculated energy band structures using the  hybrid HSE functional~\cite{HSE} for CuI and In$_2$Se$_3$ isolated monolayers (\textbf{a}) and their heterostructure in the two polarization states (\textbf{b}). For the isolated monolayers also the DFT-PBE energy bands are reported (dashed lines). In all panels color denotes the layer weight to the corresponding eigenstates as illustrated by the colorbar. For the heterostructure, we find that HSE calculations predict the same qualitative picture as DFT-PBE, with In$_2$Se$_3$/CuI behaving as a ferroelectric QSHI that has a topologically trivial gap in one polarization state (left in panel \textbf{b}) and non-trivial in the other (right). Quantitative differences are highlighted in Supplementary Table \ref{tab:beyond}.}
    \label{fig:hse}
\end{figure*}

\textit{Hybrid functionals}: We adopt the HSE functional~\cite{HSE} and perform self-consistent calculations including SOC (see Methods) for the two isolated monolayers and for the In$_2$Se$_3$/CuI heterostructure in the topological and trivial phase. As shown in Supplementary Fig.~\ref{fig:hse}a and Supplementary Table \ref{tab:beyond}, the bottom of the conduction band in In$_2$Se$_3$, $E_{\rm c}$, is only marginally affected with respect to DFT-PBE simulations, while the top of the CuI valence band is significantly shifted to more negative energies, as expected. The band alignment between the isolated materials is further affected by an increase in the vacuum level difference across the ferroelectric layer. We thus have that at the HSE level the condition i) $|E_{\rm c}-E_{\rm v}| < \Delta\phi/2$ is safely satisfied, even more robustly than in DFT-PBE, and should be resilient against the effects of an interface dipole. To test explicitly this expectation, we have performed fully self-consistent calculations of the heterostructure in both polarization states. As shown in Supplementary Fig.~\ref{fig:hse}b, while in the $\uparrow$ polarization there is trivial gap between CuI valence band maximum and In$_2$Se$_3$ conduction band bottom (although larger than in DFT-PBE), for the opposite $\downarrow$ polarization the energy bands are inverted and a gap opens at their intersections away from $\Gamma$ (see zoom inset). We thus find a qualitatively identical picture with respect to DFT-PBE, with a topological phase transition between the two polarization states, although with quantitatively different values of the band inversion and band gap (see Supplementary Table \ref{tab:beyond}). We thus have that also at the HSE level  the In$_2$Se$_3$/CuI heterostructure is confirmed to be a ferroelectric QSHI.

\textit{G$_0$W$_0$}: We perform one-shot G$_0$W$_0$ calculations with SOC for the two isolated monolayers (details in Methods).  The band gap of the two isolated monolayers is larger compared to the HSE prediction, namely 1.5 eV for In$_2$Se$_3$ (1.29 in HSE) and 3.6 eV for CuI (2.9 in HSE), as typically expected. For both In$_2$Se$_3$ and CuI, G$_0$W$_0$ predicts a small (of the order of few tens of meV) quasiparticle correction to the bottom of the conduction band while the top of the valence band is substantially lowered in energy, by 1.8 eV in CuI and 0.8 eV in In$_2$Se$_3$. Since in G$_0$W$_0$ the ground state density is not computed self-consistently but remains the same as the underlying starting-point calculation (here DFT-PBE), we thus need to take the same vacuum level difference $\Delta\phi = 1.12$~eV as in DFT-PBE calculations. Thus, at the G$_0$W$_0$ level the condition $\Delta\phi/2  > |E_{\rm c}-E_{\rm v}|$ is not met  ($E_{\rm c}-E_{\rm v}=1.4$ eV) and the heterostructure is not expected to exhibit a band inversion, although the accuracy of this prediction might be limited by the non-self-consistent, perturbative nature of the G$_0$W$_0$ method. For the same reason, we have not further investigated the effect of many-body corrections to the heterostructure in the two polarization states as this method would not be able to properly capture self-consistent effects like the charge transfer and the interface dipole discussed in the main text, and thus it cannot be employed to simulate accurately the heterostructure. We stress that the modifications of the non-local dielectric screening due to the interface could affect strongly the calculation of the self-energy, even beyond the effect of an interface dipole. That is why we consider the G$_0$W$_0$ band alignments obtained with the two isolated materials not particularly accurate in this specific context, while the self-consistent HSE approach is more appropriate.

These results suggest that particular care must be taken in future screening studies that rely on the band alignment between different 2D materials, with the importance of adopting computationally expensive beyond-DFT methods to estimate accurately the work function of the isolated 2D semiconductors and the additional caveat associated with the  need to include self-consistent effects associated with charge transfer and interface dipole in predicting the properties of the heterostructure. We remark that in an experimental setting several additional aspects come into play, such as the effect of substrate or encapsulation on the interlayer distance; those can potentially be stronger than the accuracy of the methods employed in calculating spectral properties and make the direct comparison with experiments even more challenging.

\subsection{Tight-binding model}
Here we discuss the details of the tight-binding (TB) model that we introduced in the main text to describe the topological physics of In$_2$Se$_3$/CuI heterostructures. 
The simple TB model is inspired by the three maximally-localized Wannier functions (MLWFs) \cite{wannier_review_2012} that are obtained by considering the two (four with spin-orbit coupling, SOC) highest-occupied and the single (two with SOC) lowest unoccupied bands of the In$_2$Se$_3$/CuI heterostructure. While the corresponding MLWFs are rather delocalized and exhibit a complex shape, we approximate them with $s$ and $p$ orbitals, as they transform under the same representation of the symmetry group. In fact, the general features of the corresponding band structure  do not depend on the precise shape the orbitals, as long as they are composed by one orbital that is invariant under the $C_{\rm 3v}$ symmetry (that we choose as an $s$-like orbital) and two orbitals that transform with the two-dimensional representation (that we choose as $p_x$-,$p_y$-like orbitals). The simplified spatial dependence of the atomic-like orbitals allows to construct the TB model using a Slater-Koster  \cite{SlaterKoster_prb_1954}approach and limit the number of parameters in the model. Indeed, we note that the extra symmetries of the atomic-like orbitals does not allow for some hopping terms that are instead present in the full MLWF Hamiltonian, and that contribute to some specific features in the band structure that are not relevant for our discussion.

The system is hexagonal, the direct lattice vectors in the primitive cell have the same magnitude $a$ and read:
\begin{equation}
    \mathbf{a_1} = (a,0,0)\text{,}
    \qquad 
    \mathbf{a_2} = \left(-\frac{a}{2},\frac{\sqrt{3}a}{2},0\right)\text{.}
\end{equation}
We consider as TB basis the following orbitals:
\begin{equation}
\ket{s_{\uparrow}^{\rm In_2Se_3}},\ket{s_{\downarrow}^{\rm In_2Se_3}},\ket{p_{x,\uparrow}^{\rm CuI}},\ket{p_{x,\downarrow}^{\rm CuI}}, \ket{p_{y,\uparrow}^{\rm CuI}},\ket{p_{y,\downarrow}^{\rm CuI}}
\end{equation}
where all the orbitals are centered at the origin in the $xy$-plane, while along the vertical direction they are displaced by a distance $d$ so that they are respectively localized either on In$_2$Se$_3$ or CuI as shown in Supplementary Fig.~\ref{SM_Fig1}.

\begin{figure*}[b]
    \includegraphics[width=0.15\textwidth]{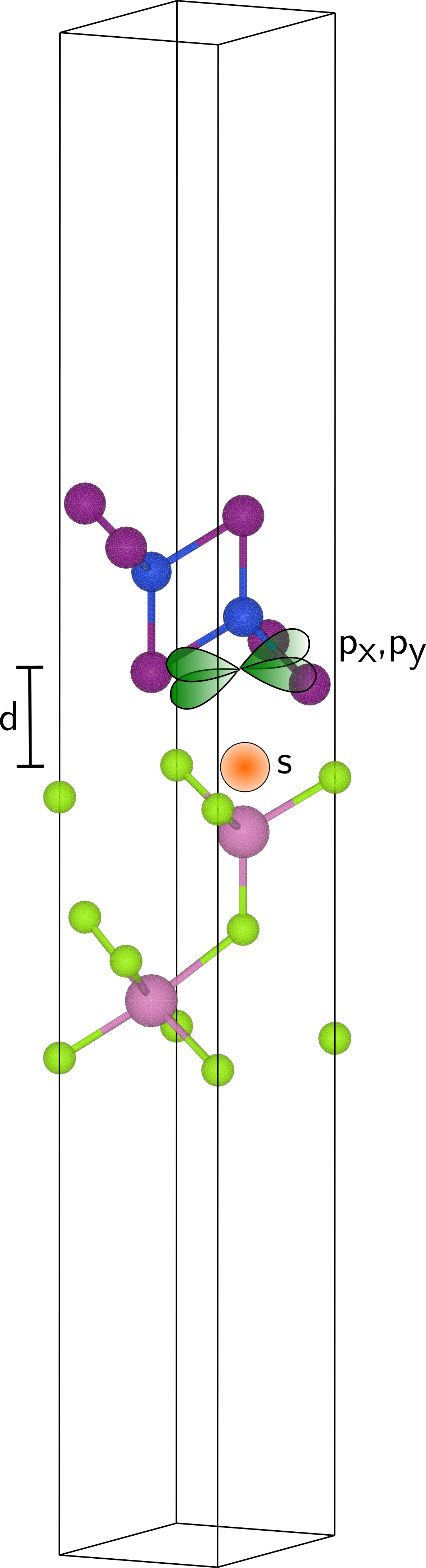}
    \caption{The tight-binding model is composed by $s$ and $p_x,p_y$ orbitals, drawn in figure, centered on In$_2$Se$_3$ and CuI respectively. The orbitals are vertically aligned and separated by an interlayer distance $d$.}
    \label{SM_Fig1}
\end{figure*}

In order to simplify the notation, we write the full TB Hamiltonian $H$ as  a sum of a spin-independent term $H_0$ and the spin-orbit coupling term $\lambda_{\rm SOC} H_{\rm SOC}$:
\begin{equation}
    H = H_0 + \lambda_{\rm SOC} H_{\rm SOC}.
\end{equation}
We start to construct the Hamiltonian $H_0$ with the Slater-Koster approach \cite{SlaterKoster_prb_1954}, by considering a real nearest-neighbour hopping term $V_{ss\sigma}$ for $s$-orbitals, while for $p_x$-orbitals and $p_y$-orbitals we write the nearest-neighbor hopping in terms of the $V_{pp\sigma}$ and $V_{pp\pi}$ orbitals depending on the orientation \cite{SlaterKoster_prb_1954}. For simplicity we drop the spin variable, meaning that the terms are identical and diagonal in the spin degree of freedom. Taking into account the relative orientation between the orbitals \cite{SlaterKoster_prb_1954}, we obtain the following nearest-neighbor matrix elements:
\begin{eqnarray}
    \braket{s^{\rm In_2Se_3}|H|s^{\rm In_2Se_3}(\mathbf{a_1})} &=& V_{ss\sigma}, \\
    \braket{s^{\rm In_2Se_3}|H|s^{\rm In_2Se_3}(\mathbf{a_2})} &=& V_{ss\sigma}, \\
    \braket{s^{\rm In_2Se_3}|H|s^{\rm In_2Se_3}(\mathbf{a_1}+\mathbf{a_2})} &=& V_{ss\sigma}, \\
    \braket{p_{x}^{\rm CuI}|H|p_{x}^{\rm CuI}(\mathbf{a_1})} & = & V_{pp\sigma},\\
    \braket{p_{x}^{\rm CuI}|H|p_{x}^{\rm CuI}(\mathbf{a_2})}& = & \frac{V_{pp\sigma}}{4}+\frac{3V_{pp\pi}}{4},\\
    \braket{p_{x}^{\rm CuI}|H|p_{x}^{\rm CuI}(\mathbf{a_1}+\mathbf{a_2})}& = & \frac{V_{pp\sigma}}{4}+\frac{3V_{pp\pi}}{4},\\
    \braket{p_{y}^{\rm CuI}|H|p_{y}^{\rm CuI}(\mathbf{a_1})} & = & V_{pp\pi},\\
    \braket{p_{y}^{\rm CuI}|H|p_{y}^{\rm CuI}(\mathbf{a_2})}& = & \frac{V_{pp\pi}}{4}+\frac{3V_{pp\sigma}}{4},\\
    \braket{p_{y}^{\rm CuI}|H|p_{y}^{\rm CuI}(\mathbf{a_1}+\mathbf{a_2})}& = & \frac{V_{pp\pi}}{4}+\frac{3V_{pp\sigma}}{4},\\
    \braket{p_{x}^{\rm CuI}|H|p_{y}^{\rm CuI}(\mathbf{a_2})}& = & \frac{\sqrt{3}}{4}(V_{pp\pi}-V_{pp\sigma}),\\
    \braket{p_{x}^{\rm CuI}|H|p_{y}^{\rm CuI}(\mathbf{a_1}+\mathbf{a_2})}& = & \frac{\sqrt{3}}{4}(V_{pp\sigma}-V_{pp\pi}),\\
    \braket{p_{y}^{\rm CuI}|H|p_{x}^{\rm CuI}(\mathbf{a_2})}& = & \frac{\sqrt{3}}{4}(V_{pp\pi}-V_{pp\sigma}),\\
    \braket{p_{y}^{\rm CuI}|H|p_{x}^{\rm CuI}(\mathbf{a_1}+\mathbf{a_2})}& = & \frac{\sqrt{3}}{4}(V_{pp\sigma}-V_{pp\pi}),
\end{eqnarray}
where the position variable (e.g. $\mathbf{a}_1$) represents the cell where the orbital is centered (absent if centered in the home cell).
The energy offset between the orbitals on the two layers is accounted for by a diagonal on-site term:
\begin{equation}
    \braket{s^{\rm In_2Se_3}|H|s^{\rm In_2Se_3}}-\braket{p_{x,y}^{\rm CuI}|H|p_{x,y}^{\rm CuI}} = \Delta.
\end{equation}
Now we include the interlayer interaction through a hopping term between $s$ and $p$ orbitals:
\begin{eqnarray}
    \braket{s^{\rm In_2Se_3}|H|p_{x}^{\rm CuI}(\mathbf{a_1})} &=& \tilde{V},  \\
    \braket{s^{\rm In_2Se_3}|H|p_{x}^{\rm CuI}(\mathbf{a_2})} &=& -\frac{\tilde{V} }{2},\\
    \braket{s^{\rm In_2Se_3}|H|p_{x}^{\rm CuI}(\mathbf{a_1}+\mathbf{a_2})} &=& \frac{\tilde{V}}{2},\\
    \braket{p_{x}^{\rm CuI}|H|s^{\rm In_2Se_3}(\mathbf{a_1})} &=& -\tilde{V},  \\
    \braket{p_{x}^{\rm CuI}|H|s^{\rm In_2Se_3}(\mathbf{a_2})} &=& \frac{\tilde{V} }{2}, \\
    \braket{p_{x}^{\rm CuI}|H|s^{\rm In_2Se_3}(\mathbf{a_1}+\mathbf{a_2})} &=&-\frac{\tilde{V} }{2},\\
    \braket{s^{\rm In_2Se_3}|H|p_{y}^{\rm CuI}(\mathbf{a_2})} &=& \frac{\sqrt{3}\tilde{V} }{2}, \\
    \braket{s^{\rm In_2Se_3}|H|p_{y}^{\rm CuI}(\mathbf{a_1}+\mathbf{a_2})} &=& \frac{\sqrt{3}\tilde{V} }{2},\\
    \braket{p_{y}^{\rm CuI}|H|s^{\rm In_2Se_3}(\mathbf{a_2})} &=& -\frac{\sqrt{3}\tilde{V} }{2}, \\
    \braket{p_{y}^{\rm CuI}|H|s^{\rm In_2Se_3}(\mathbf{a_1}+\mathbf{a_2})} &=& -\frac{\sqrt{3}\tilde{V} }{2}.\\
\end{eqnarray}
The coupling $\tilde{V}$ can be related to the Slater-Koster interaction integral $V_{sp\sigma}$  through $ \tilde{V} = V_{sp\sigma}\cos(\phi)$, where $\phi$ is angle between the plane of one monolayer and the vector that connects the two orbitals, that is
$\cos(\phi) = {a}/{\sqrt{a^2+d^2}}$.

Finally we include the on-site spin-orbit coupling through the term $L_z\sigma_z$ that connects $p_x$ and $p_y$ orbitals:
\begin{eqnarray}
    \braket{p_{x\uparrow}^{\rm CuI}|H_{\rm SOC}|p_{y\uparrow}^{\rm CuI}}& = & -i\lambda_{\rm SOC},\\
    \braket{p_{x\downarrow}^{\rm CuI}|H_{\rm SOC}|p_{y\downarrow}^{\rm CuI}}& = & i\lambda_{\rm SOC}.
\end{eqnarray}

\begin{table}[h!]
\caption{\label{tab:params} Values of tight-binding parameters (in meV) adopted for the calculations shown in the main text.} 
\begin{tabularx}{0.6\linewidth}{>{\centering\arraybackslash}X >{\centering\arraybackslash}X >{\centering\arraybackslash}X >{\centering\arraybackslash}X >{\centering\arraybackslash}X >{\centering\arraybackslash}X}
\toprule
$\Delta$ & $V_{ss\sigma}$ & $V_{pp\sigma}$ & $V_{pp\pi}$   & $\tilde V$ & $\lambda_{\rm \rm SOC}$\\
\midrule 
900  & $-80$ & 200 & $-40$ & 20 & 50\\
\bottomrule
\end{tabularx}
\end{table}

\end{document}